\documentclass[conference]{IEEEtran}
\IEEEoverridecommandlockouts

\usepackage{cite}
\usepackage{amsmath,amssymb,amsfonts}
\usepackage{graphicx}
\usepackage{textcomp}
\usepackage{xcolor}

\usepackage{kotex}
\usepackage{url}
\usepackage{subcaption}
\usepackage{graphicx}
\usepackage{comment}
\usepackage{booktabs}
\usepackage[linesnumbered,ruled,vlined]{algorithm2e}
\usepackage{float}
\usepackage{tabularx}
\usepackage{hyperref}
\usepackage{tikz}
\usepackage{amsmath,amssymb,amsfonts}
\usepackage{textcomp}
\usepackage[table, dvipsnames]{xcolor}
\usepackage[normalem]{ulem}
\usepackage{titlecaps}

\usepackage{url}
\usepackage{hyperref}
\hypersetup{hidelinks,colorlinks=true,linkcolor=green!80!black,citecolor=red!70!black,urlcolor=blue!70!black}
\usepackage{bm}

\usepackage{multirow}
\usepackage{array}
\usepackage{pifont}
\usepackage{mathtools}
\usepackage{array}

\usepackage{tikz}

\usepackage[most]{tcolorbox} 

\newcommand{\squishlist}{
\begin{list}{$\bullet$}
	{ \setlength{\itemsep}{0pt}      \setlength{\parsep}{-0pt}
		\setlength{\topsep}{4pt}       \setlength{\partopsep}{0pt}
		\setlength{\listparindent}{-2pt}
		\setlength{\itemindent}{-5pt}
		\setlength{\leftmargin}{1em} \setlength{\labelwidth}{0em}
		\setlength{\labelsep}{0.5em} } }
\newcommand{\squishend}{
\end{list}  }

\newcommand{\cred}{\textcolor{red}}

\newcommand{\sol}{\textsc{Resystance}}
\newcommand{\solk}{\textsc{Resystance-k}}

\usepackage{etoolbox}

\usepackage{capt-of}

\begin{document}

\title{\sol{}: Unleashing Hidden Performance of Compaction in LSM-trees via eBPF}

\makeatletter
\newcommand{\linebreakand}{%
  \end{tabular}\hfill\par
  \vspace{0.8em}%
  \hfill\begin{tabular}[t]{c}%
}
\makeatother

\author{\IEEEauthorblockN{Hongsu Byun\IEEEauthorrefmark{1}, Seungjae Lee\IEEEauthorrefmark{1}, Honghyeon Yoo, Myoungjoon Kim, Sungyong Park\IEEEauthorrefmark{2}
\thanks{\IEEEauthorrefmark{1}These authors are first co-authors and have contributed equally.
}
\thanks{\IEEEauthorrefmark{2}S. Park is the corresponding author.
}
}
\IEEEauthorblockA{\textit{Sogang University},
Seoul, Republic of  Korea\\
\{byhs, ctaaone, yhh, myoungjoon, parksy\}@sogang.ac.kr
}
}

\maketitle

\begin{abstract}
The development of high-speed storage devices such as NVMe SSDs has shifted the primary I/O bottleneck from hardware to software. Modern database systems also rely on kernel-based I/O paths, where frequent system call invocations and kernel–user space transitions lead to relatively large overheads and performance degradation. This issue is particularly pronounced in Log-Structured Merge-tree (LSM-tree)–based NoSQL databases. We identified that, in particular, the background compaction process generates a large number of read system calls, causing significant overhead.
To address this problem, we propose \sol{}, which leverages eBPF and io\_uring to free compaction from system calls and unlock hidden performance potential. \sol{} improves disk I/O efficiency during read operations via io\_uring and significantly reduces software stack overhead by handling compaction directly inside the kernel through eBPF. Moreover, \sol{} minimizes user–kernel transitions by offloading key I/O routines into the kernel without modifying the LSM-tree structure or compaction algorithm.
\sol{} was extensively evaluated using db\_bench, YCSB, and OLTP workloads. Compared to baseline RocksDB, it reduced the average number of system call invocations during compaction by 99\% and shortened compaction time by 50\%. Consequently, in write-intensive workloads, \sol{} improved throughput by up to 75\% and reduced the p99 latency by 40\%.
\end{abstract}


\vspace{-10pt}
\setlength{\textfloatsep}{5pt plus 1.0pt minus 2.0pt}
\section{Introduction}
\noindent\textbf{Shift of the I/O Bottleneck.}
The emergence of ultra-fast storage devices such as NVMe SSDs~\cite{Samsung_9100PRO_2025,Intel_Optane_P5800X_2021,Micron_7600_NVMe_2025} has fundamentally changed the nature of system performance bottlenecks. In the past, limited disk bandwidth and high latency were the dominant constraints, whereas modern storage now delivers several GB/s of bandwidth and microsecond-level latency~\cite{he2023database,KIOXIA_CD8P_R_2023,SSSTC_EJ5_U2_2024}. Consequently, the performance ceiling of I/O operations no longer resides in the hardware layer. The I/O bottleneck has thus shifted from hardware to software~\cite{zhong2022xrp}. Each I/O request incurs multiple system calls and frequent context switches between kernel and user space, wasting CPU cycles and preventing full utilization of high-performance devices. This shift affects not only kernel-level operations but also data-intensive applications that depend on them. Database systems are no exception: due to multi-layered I/O paths for buffer management, logging, and file access, they are particularly vulnerable to software-induced overheads. As a result, the primary determinant of database performance has moved from hardware limitations to system call overheads arising at the OS–DBMS boundary.

\noindent\textbf{System Call Bottlenecks in LSM-trees.}
This software-induced overhead is widely observed across databases but is especially pronounced in NoSQL systems. Representative examples such as RocksDB~\cite{rocksdb}, LevelDB~\cite{levelDB}, and Cassandra~\cite{lakshman2010cassandra} employ the Log-Structured Merge-tree (LSM-tree)~\cite{o1996log} as their storage engine. The LSM-tree architecture maximizes write efficiency by buffering writes in memory and periodically flushing them to disk in batches, while continuously merging (compacting) data in the background~\cite{rocksdb-compaction,sarkar2022constructing}. Although this design provides high write throughput, the flush and compaction operations trigger intensive file accesses and consequently frequent system calls.

Our profiling shows that compaction dominates the total number of system calls within the LSM-tree engine, with \texttt{pread()} accounting for the majority. This is because compaction requires kernel-mediated I/O for each of the thousands of SSTable blocks it repeatedly reads and merges. As a result, compaction becomes the most severe system call bottleneck in the I/O path of LSM-tree–based databases. A detailed analysis is presented in Section~\ref{sec:back}.

\noindent\textbf{Constraints of Kernel Bypass.}
Several efforts have aimed to mitigate such kernel overheads~\cite{196386,lee2019asynchronous,yang2017spdk,spdk,dpdk,zhang2019m}. 
In particular, kernel-bypass frameworks such as SPDK (Storage Performance Development Kit)~\cite{spdk} eliminate system call overhead by allowing user-space applications to directly access NVMe devices, achieving extremely high I/O parallelism.
However, this approach presents significant practical limitations for database systems. 
Because SPDK bypasses the OS file system, adopting it requires abandoning traditional file-based data management and redesigning the entire I/O subsystem. Such radical architectural changes introduce substantial implementation complexity, maintenance costs, and incompatibility with standard OS interfaces. Hence, while kernel bypass is theoretically efficient, it remains impractical for mature database systems that must preserve their existing architectures.

\noindent\textbf{The Case for eBPF in LSM-tree Compaction. }
To address these bottlenecks without the drawbacks of kernel bypass, we look toward the extended Berkeley Packet Filter (eBPF)~\cite{fleming2017_ebpf}. As an in-kernel programmable framework, eBPF is suited for optimizing LSM-tree compaction for three reasons. First, \textit{Safety}: eBPF’s verifier ensures that injected logic cannot crash the kernel, preserving database reliability. Second, \textit{Efficiency}: It allows executing user-defined logic directly in the kernel context, eliminating  context switches and data copying. Third, \textit{Practicality}: It enables dynamic acceleration without modifying the kernel source code or the underlying file system, offering a non-intrusive path to performance optimization.

\noindent\textbf{Challenges.}
However, the design of compaction using eBPF presents the following challenges:
\squishlist
\item eBPF imposes strict constraints on instruction count, loop structure, and stack usage. Designing compaction logic within these constraints is highly complex.
\item Compaction in eBPF interacts with external data and kernel components. Since the verifier checks only internal eBPF logic, potential infinite loops or kernel panics arising from such interactions cannot be fully prevented.
\item This solution must maintain kernel maintainability and ensure portability across various LSM-tree-based systems, while avoiding intrusive changes to the kernel or system-specific deep redesigns of the LSM-tree engine.
\squishend

\noindent\textbf{\sol{}: System Call Liberation for eBPF-based Compaction.}
To address these challenges, we propose \sol{}, an eBPF-based framework that frees LSM-tree compaction from system call overhead. The name “Resystance,” inspired by “Resistance,” reflects its goal of resisting the overhead imposed by system calls and liberating compaction from their dependence. Without modifying the structure or compaction algorithms of existing LSM-tree engines such as RocksDB, and without requiring additional hardware accelerators or NVM, \sol{} combines eBPF~\cite{fleming2017_ebpf} and io\_uring~\cite{man7-iouring} to migrate the core I/O routines of compaction into the kernel. By minimizing user–kernel boundary crossings, it eliminates unnecessary system calls along the data path and achieves performance improvements purely through software stack optimization.

To the best of our knowledge, \sol{} is the first system by solely eliminating software stack overhead. \sol{} liberates LSM-tree compaction from the constraints of system calls and enhances database I/O efficiency.

\noindent\textbf{Contributions.}
The contributions of this paper are summarized as follows:
\squishlist
\item We identify compaction as the dominant source of system call overhead in LSM-tree–based NoSQL databases.
\item We propose \sol{}, which offloads the core I/O logic of compaction into the kernel by integrating eBPF and io\_uring.
\item \sol{} significantly reduces system call overhead and improves disk I/O efficiency without any hardware modification or LSM-tree redesign.
\squishend

\sol{} is implemented on a representative LSM-tree–based key-value store, RocksDB v10.1.3, and extensively evaluated using db\_bench, YCSB benchmarks, and OLTP workloads. Compared to the baseline RocksDB, it reduces the average number of system calls during compaction by 99\% and shortens compaction latency by 50\%. Specifically, in the db\_bench FillRandom workload, \sol{} improves throughput by up to 75\% and achieves 40\% lower p99 latency. Furthermore, \sol{} achieves 22\% higher throughput than baseline RocksDB in mixed read-write OLTP workloads.

\begin{figure}[!t]
	\centering
    \includegraphics[width=0.95\linewidth]{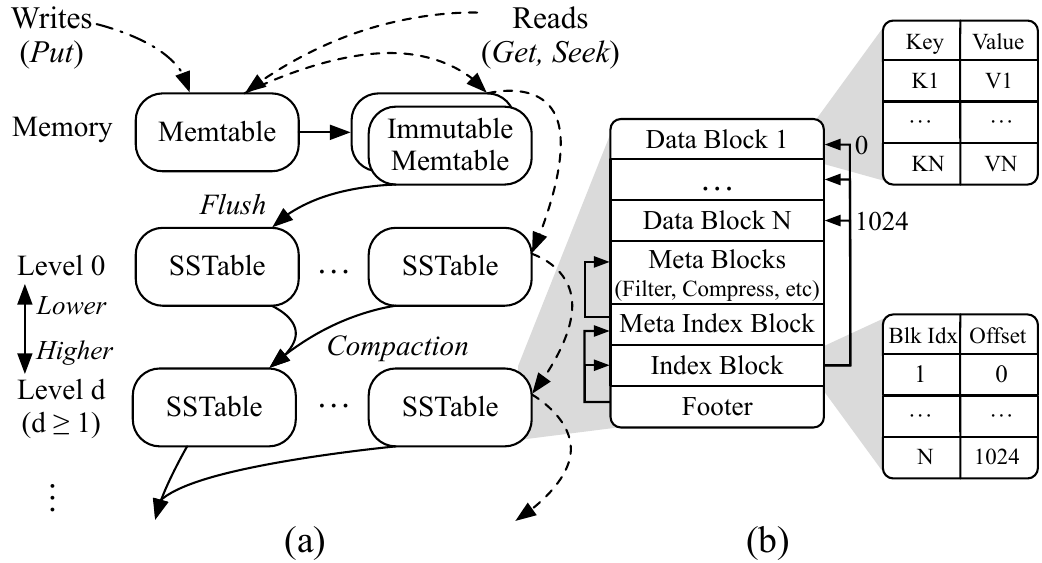}
    \caption{An Architecture of LSM-tree.
	}
	\label{fig:back_lsm}
\end{figure}

\section{Background and Problem Definition}
\label{sec:back}
This section describes the background of Log-Structured Merge-trees (LSM-trees) and analyzes the performance bottlenecks and limitations observed in today’s LSM-tree–based NoSQL databases.

\subsection{Log-structured Merge-tree}
Figure~\ref{fig:back_lsm}(a) illustrates the structure of an LSM-tree. An LSM-tree buffers key–value \textit{writes} in a memory structure called the Memtable and periodically performs a \textit{flush} to persist the data as an on-disk structure, the SSTable. This batching effect converts frequent random writes into large sequential writes, providing high write throughput. Newly flushed SSTables are first written to level~0, and when the number of SSTables in level~0 exceeds a predefined threshold, a \textit{compaction} process merges and sorts them, producing a new SSTable in the next level. Each level maintains its own compaction threshold, and the capacity of each level increases exponentially with the level number.

A \textit{read} operation first searches the Memtable and then scans SSTables sequentially from lower to higher levels. When redundant or unsorted keys accumulate, the lookup time increases; thus, compaction—which removes duplicates and reorganizes key ranges—is essential to maintaining fast reads. When compaction is delayed, reads must traverse numerous SSTables, causing a steep performance drop. Consequently, LSM-trees cannot sustain read performance during compaction backlogs, and foreground writes are often paused to prevent further degradation, leading to a \textit{write stall}~\cite{yao2020matrixkv, ding2022trianglekv, wang2023rbc, yu2023adoc, writestall}.

\noindent\textbf{Importance of Compaction Performance.}
Compaction affects both write and read performance in LSM-trees, making it a critical determinant of overall system efficiency. As such, improving compaction performance has been a central research topic, ranging from hardware acceleration using FPGA, GPU, or DPU devices to algorithmic optimizations and new compaction policies. Section~\ref{sec:related} reviews these studies in detail.

\subsection{Bottleneck of Software Stack}
Recent work~\cite{zhong2022xrp} shows that even simple file I/O calls such as \texttt{read()} or \texttt{write()} on modern NVMe SSDs suffer greater latency from the software stack than from the hardware itself. As SSD latency and internal parallelism improve dramatically, the relative cost of kernel-level operations—data copying, buffer management, and context switching—dominates total I/O latency. In other words, the bottleneck has shifted from slow storage hardware to the software layers of the operating system and its system call interface.

\begin{table}[t]
    \centering 
    \caption{Latency breakdown of \texttt{pread()} in LSM-trees.}
    \vspace{-4pt}
    \renewcommand{\arraystretch}{1.1}
    \resizebox{0.4\textwidth}{!}{
    {
\begin{tabular}{c|lrr}
\toprule
 & File System  & 110.4\,µs & 61.2\% \\ 
Software & Block Layer  & 5.9\,µs & 3.8\% \\ 
& NVMe Driver  & 6.3\,µs & 3.5\% \\ \midrule
Hardware & Storage Device  & 56.8\,µs & 31.5\% \\ \midrule
& Total & 180.4\,µs & 100\% \\ \bottomrule
\end{tabular}
    }
    }
    \label{tbl:motiv_lsm_read_time_break}
\end{table}

\vspace{6pt}
\begin{table}[t]
    \centering 
    \caption{Average number of system calls per operation in RocksDB.}
    \renewcommand{\arraystretch}{1.3}
    \vspace{-4pt}
    \resizebox{0.47\textwidth}{!}{
    {
\begin{tabular}{ccccccc}
\toprule
\textbf{} & \textbf{Put} & \textbf{Get} & \textbf{Seek} & \textbf{Next} & \textbf{Flush} & \textbf{Compaction} \\ \midrule
\textbf{Syscall/Op} & 0 & 0.5 & 0.5 & 0.01 & 107 & \cred{66{,}983} \\ \bottomrule
\end{tabular}
    }
    }
    \label{tbl:motiv_syscall_per_op}
\end{table}

\vspace{6pt}
\begingroup
\setlength{\textfloatsep}{0pt} 
\begin{table}[!t]
    \centering 
    \caption{System call distribution during compaction.}
    \vspace{-4pt}
    \renewcommand{\arraystretch}{1.3}
    \resizebox{0.47\textwidth}{!}{
    {
\begin{tabular}{cccccc}
\toprule
\textbf{} & \textbf{write} & \textbf{fsync} & \textbf{pread} & \textbf{unlink} & \textbf{others} \\ \midrule
\textbf{Ratio(\%)} & 0.38 & 0.005 & \cred{96.09} & 0.01 & 3.52 \\ \bottomrule
\end{tabular}
    }
    }
    \label{tbl:motiv_syscall_ratio}
     \vspace{0pt}
\end{table}
\endgroup

This observation extends directly to database systems. Table~\ref{tbl:motiv_lsm_read_time_break} shows the latency breakdown of a \texttt{pread()} during a \texttt{Get} operation in RocksDB, a representative LSM-tree–based key-value store. The software stack accounts for 68.5\% of the total latency. Both user queries and internal operations such as flush and compaction rely heavily on kernel-mediated system calls. Consequently, modern databases are no longer limited by storage hardware but rather by the software overhead of system calls within the kernel I/O path.

\begingroup
\setlength{\textfloatsep}{0pt} 
\begin{figure}[t]
	\centering
    \includegraphics[width=0.65\linewidth]{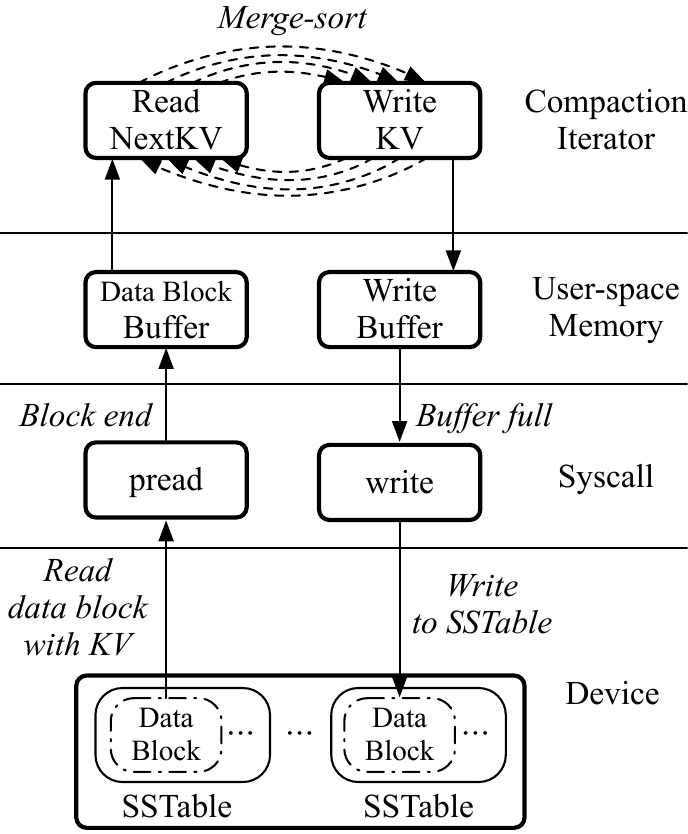}
    \caption{Operation of block-based compaction in LSM-trees.}
	\label{fig:back_compaction}
\end{figure}
\endgroup

\subsection{Analysis of LSM-tree System Calls}
We quantitatively analyze which operations within the LSM-tree generate most system calls using the db\_bench under the FillRandom and MixGraph workloads. The MixGraph workload~\cite{cao2020characterizing}, modeled after Facebook’s RocksDB usage pattern, follows the default ratio of 83\% \texttt{Get}, 14\% \texttt{Put}, and 13\% \texttt{Seek}. Experimental details are provided in Section~\ref{sec:eval}. Table~\ref{tbl:motiv_syscall_per_op} compares the average number of system calls per operation, revealing that \textit{compaction} overwhelmingly dominates. This is because compaction repeatedly reads and writes a large number of SSTable files. Hence, compaction is not only a data reorganization task but also the most system-call–intensive operation within the LSM-tree engine.

To further understand this behavior, we examine the system call composition during compaction. As shown in Table~\ref{tbl:motiv_syscall_ratio}, \texttt{pread()} accounts for the vast majority of calls. This is due to its block-based design, where SSTables are read block by block and key–value pairs are merged and rewritten. Figure~\ref{fig:back_compaction} illustrates the detailed operation of compaction: the compaction iterator repeatedly reads (\texttt{ReadNextKV}) and writes (\texttt{WriteKV}) key–value pairs in a merge-sort fashion. For each required block, a \texttt{pread()} is issued. Assuming a 64\,MB SSTable and 4\,KB data blocks, each SSTable contains $2^{14}$ blocks, leading to an enormous number of \texttt{pread()} calls. This block-based structure is thus the principal source of software-stack overhead.

\subsection{Why LSM-trees Adopt Block-Based SSTables}
Why do LSM-trees adopt block-based SSTable formats? The reason lies in achieving both write and read efficiency.

On the write side, LSM-trees convert random writes into sequential writes to maximize disk efficiency. As illustrated in Figure~\ref{fig:back_lsm}(b), SSTables are organized into multiple \textit{data blocks}. During flush or compaction, data are written sequentially across contiguous regions, which effectively utilizes the file system’s and SSD’s internal buffers and parallel channels, improving throughput and mitigating write amplification. Thus, block-based storage enables the “large sequential write” behavior that LSM-trees are designed to exploit.

On the read side, the block-based structure is equally essential. An SSTable consists of metadata blocks and data blocks; the \textit{index block} stores the offset, size, and key range of each data block. A lookup first queries the index block and then accesses only the corresponding data block, reducing unnecessary I/O and improving efficiency. This multi-level lookup reduces metadata size and enhances caching effectiveness. Therefore, block-based SSTables provide both sequentialization for writes and locality for reads, forming the structural foundation of LSM-tree performance.

\section{Opportunities for Reducing System Calls}
This section introduces the frameworks that enable reducing system calls in compaction and explains the constraints of existing kernel-bypass approaches.

\subsection{Extended Berkeley Packet Filter}
eBPF (extended Berkeley Packet Filter) enables execution of verified bytecode programs within the Linux kernel by dynamically attaching them to various hook points (e.g., kprobes/tracepoints, uprobes, XDP, and etc.). Each program is statically verified by the kernel verifier for memory safety, guaranteed termination, and helper call restrictions, and then JIT-compiled for execution at the attached hook point in an event-driven manner. Through eBPF maps (shared key–value state) and helper functions, eBPF has been widely applied to networking (packet filtering and load balancing), observability (profiling and tracing), and security (policy enforcement and syscall monitoring). Features such as BTF and CO-RE (Compile Once–Run Everywhere) allow a single eBPF bytecode to run seamlessly across different kernel versions.

eBPF allows user-defined logic to execute at the kernel level. Specifically, by offloading selected application functionalities from the I/O path into eBPF programs, multiple system call effects can be achieved within a single kernel invocation. Moreover, data can be preprocessed in the kernel before being passed to user space, reducing user memory demand and improving working-set efficiency. This approach removes or amortizes context switches and user–kernel boundary crossings, improving cache locality while reducing TLB pressure and memory bandwidth waste. Although these benefits must operate under verifier constraints and kernel stability considerations, eBPF provides practical performance opportunities for optimizing I/O paths.

\subsection{IO\_uring}
io\_uring is a modern asynchronous I/O interface that shares submission and completion rings (SQ/CQ) between user and kernel spaces via \texttt{mmap}, thereby eliminating per-I/O system calls and enabling syscall amortization through batched operations. Applications fill submission queue entries (SQEs) with operations such as \texttt{read}, \texttt{write}, \texttt{poll}, and \texttt{fsync}, while the kernel asynchronously returns results through completion queue entries (CQEs). By sharing memory and supporting rich asynchronous primitives, io\_uring minimizes context-switch and data-copy overheads, providing an efficient execution model in which multiple I/Os are submitted and completed through a single kernel entry.

\subsection{Constraints and limitations of kernel bypass}
Similarly, approaches such as XRP, which provide a programmable fast path at the storage-driver level to process or transform requests in-driver, can shorten the syscall path but bypass the VFS and block layers. As a result, they cannot fully leverage kernel functionalities such as I/O scheduling, writeback, and readahead. This limitation, combined with file-system feature constraints, greatly increases implementation complexity. In workloads that operate over many files and directories (e.g., numerous small I/Os distributed across multiple files or per-file policies), ensuring consistency and controlling shared resources become particularly difficult. Ultimately, existing methods that rely on kernel bypass or block/file-system layer omission struggle to achieve generality, safety, and operational viability. A new approach is required—one that preserves the advantages of kernel layers while reducing user–kernel boundary overhead.

\section{Design Goals and Challenges}
This section presents the design goals and the challenges of accelerating compaction using eBPF and io\_uring.

\subsection{Goals}
\noindent\textbf{Goal \#1: Minimizing overhead via syscall reduction.}
Conventional iterator-based compaction issues a \texttt{pread()} system call for every data-block read, incurring massive syscall counts and context-switch overheads. \sol{} leverages eBPF to offload the core I/O and merge-sort phases of compaction into the kernel, thereby reducing user–kernel crossings and mitigating syscall-induced overheads.

\noindent\textbf{Goal \#2: Improving disk I/O efficiency with in-kernel asynchronous I/O.}
In the conventional path, each data-block access traverses the kernel to reach the storage device through a \texttt{pread()} call, effectively serializing I/O per call forcing synchronous I/O per block read. In contrast, because \sol{} executes compaction within the kernel, it can employ io\_uring to request many data blocks asynchronously and in parallel. This reduces kernel entries and maximizes the utilization of the intrinsic parallelism and bandwidth of high-performance NVMe SSDs.

\noindent\textbf{Goal \#3: Enabling dynamic compaction algorithms.}
Traditional compaction typically uses a fixed algorithm (e.g., a min-heap–based merge), which is difficult to adapt to workload characteristics or device conditions. \sol{} removes this limitation by allowing the compaction algorithm itself to be defined in user space and injected into the kernel using eBPF. As a result, the system can replace compaction logic at runtime—without database restarts or kernel rebuilds—and dynamically adopt the most suitable merge strategy to improve effectiveness and adaptability.

\noindent\textbf{Goal \#4: Minimizing kernel modifications.}
Excessive kernel modifications risk degrading stability and hinder real-world deployment. \sol{} therefore minimizes kernel changes to improve portability across environments. To minimize interference between the application and the operating system interfaces, \sol{} is designed to operate through existing system calls rather than introducing new ones. Moreover, to reduce the impact on the original kernel control flow, the additional functionality introduced by this work is invoked from a top-level function with a low call frequency, ensuring that modifications remain minimal and localized.

\subsection{Challenges}
\noindent\textbf{Challenge \#1: eBPF performance limits and verifier constraints.}
Although \sol{} moves part of the compaction logic into the kernel via eBPF, eBPF enforces strict limits on instruction count, loop structure, stack usage, and call depth for safety. These constraints make it difficult to implement operations such as merge-sort or iterators purely in eBPF, and the likelihood of verifier rejection increases with complexity. Furthermore, due to map accesses and helper invocations, eBPF often performs worse than native kernel code~\cite{volpert2025towards,liu2024understanding}, which can become a bottleneck for I/O- and sort-intensive compaction.

\noindent\textbf{Challenge \#2: Function partitioning between kernel and eBPF logic.}
Given these constraints, \sol{} must carefully split responsibilities between kernel code and eBPF programs. Overloading eBPF increases the risk of verifier failure and system instability; under-specifying eBPF leaves it too weak to express the required compaction behavior. \sol{} therefore needs a clear boundary and a partitioning policy that preserves sufficient flexibility while staying within verifier limits.

\noindent\textbf{Challenge \#3: Kernel stability.}
While the eBPF verifier prevents issues such as infinite loops and invalid memory accesses within eBPF programs, it does not verify interactions with external kernel code. If user-supplied logic interacts improperly with kernel resources, it may cause infinite loops or kernel panics. To protect overall system stability, \sol{} must restrict resource access, and incorporate failure isolation and recovery mechanisms.

\section{\sol{} Design}
This section describes the architecture of \sol{}. We first present an overview and then the execution workflow.

\begingroup
\setlength{\textfloatsep}{0pt}  
\begin{figure}[t]
	\centering
    \includegraphics[width=0.85\linewidth]{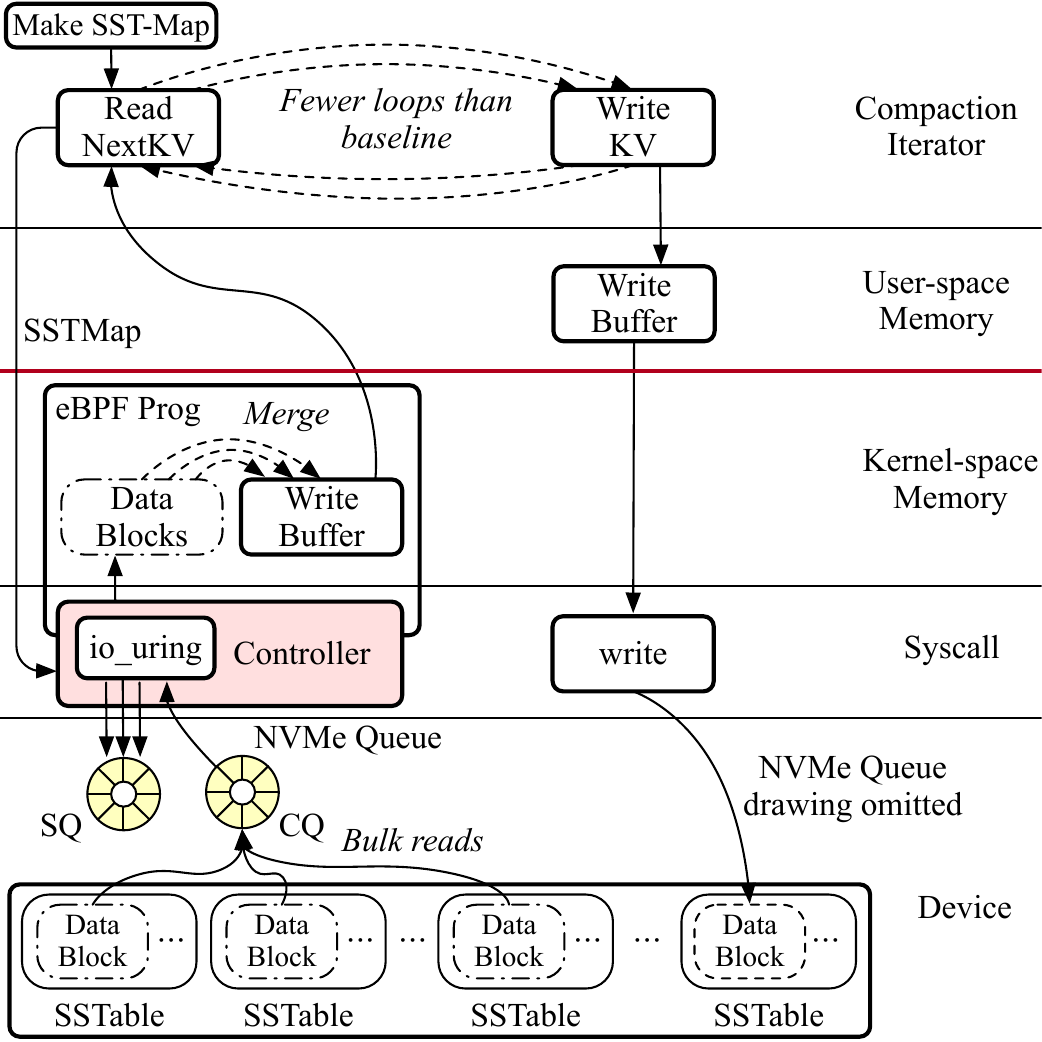}
    \caption{Overview of the \sol{} architecture.
	}
	\label{fig:design_solution_overview}
\end{figure}
\endgroup

\begingroup
\setlength{\textfloatsep}{0pt}  
\begin{figure*}[t]
	\centering
    \includegraphics[width=1\linewidth]{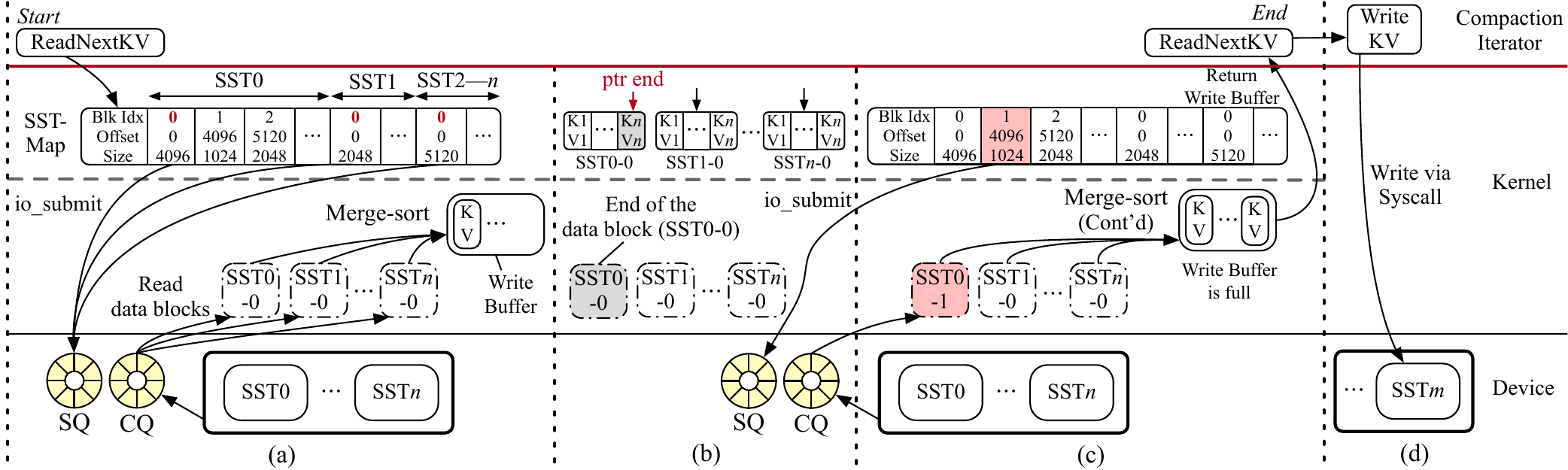}
    \caption{Detailed operation of the \sol{} controller.}
	\label{fig:design_solution_detail}
    \vspace{-10pt}
\end{figure*}
\endgroup

\subsection{Overview}
\label{sec:design_overview}
\noindent\textbf{Core Idea.}
Figure~\ref{fig:design_solution_overview} shows the layered architecture of \sol{}, which consists of three components: the controller, io\_uring, and eBPF programs. The key idea is to reduce the number of compaction-iterator invocations by (i) issuing multiple read I/Os via io\_uring within a single \texttt{ReadNextKV} call and (ii) performing merge-sort in an attached eBPF program. Because io\_uring is used, \sol{} can operate over the conventional kernel VFS and block layer, preserving generality.

\noindent\textbf{SST-Map for eBPF Program.}
To perform compaction, the kernel needs I/O descriptors—such as the target SSTable, block offsets, and block sizes—but this information is not inherently available to eBPF. \sol{} therefore constructs an \textit{SST-Map} to convey such metadata. The SST-Map is built in user–kernel shared memory (eliminating copy overhead) and is derived only from metadata of SSTables already loaded into main memory, making its construction cost negligible. Further details of the SST-Map are provided in the next subsection.

\noindent\textbf{Read and Merge Procedure using eBPF and io\_uring.}
When the compaction iterator invokes \texttt{ReadNextKV}, \sol{} issues asynchronous reads for the required data blocks through io\_uring. The eBPF program—hooked on the relevant event—then performs merge-sort over the read data and appends results to an output buffer. At this point, the eBPF program reads the key-value structure of the data block in the SSTable, allowing execution regardless of the size of the key-value pair. If the eBPF program exhausts a data block, the controller fetches the next block’s metadata from the SST-Map. To avoid burdening users with complex eBPF logic (which risks verifier rejection and stability issues), the controller provides the control skeleton optimized for sequential-read–then-merge patterns. Inside eBPF, the program only checks per-block validity and minimal progress state needed to continue compaction, reducing user implementation burden and minimizing kernel intrusion risk. The loop continues until either (i) the output buffer reaches a predefined size or (ii) there are no more blocks to read. Results are returned via user–kernel shared memory, removing copy overhead.

\noindent\textbf{Write Procedure via the Conventional Path.}
Afterward, we retain the original compaction iterator’s \texttt{WriteKV()} for write operations, which writes data blocks via the conventional user-space path.
As shown in Table~\ref{tbl:motiv_syscall_ratio}, \texttt{WriteKV()} issues significantly fewer system calls through \texttt{write()} than \texttt{pread()}, and it also handles SSTable metadata construction. As a result, re-implementing it in eBPF would yield marginal benefits while increasing verifier risk.

\noindent\textbf{Design Choices for eBPF-Based Compaction Logic.}
From a merge-centric perspective, several design choices are possible when applying eBPF to compaction: (i) using eBPF for read, merge, and write; (ii) read and merge; (iii) merge and write; and (iv) using eBPF only for merge. In \sol{}, we choose to offload only the merge functionality to an eBPF program in order to preserve kernel safety, minimize code modifications, and maximize compatibility with existing compaction logic while still benefiting from eBPF. A detailed discussion of this design choice is provided in the Section~\ref{sec:discuss}.

\subsection{Kernel Stability Maintenance}
\label{sec:design_kernel}
\noindent\textbf{eBPF Verifier Modification for Kernel Stability.}
To realize \sol{} while preserving kernel stability, we modify the eBPF verifier. First, we relax the verification instruction limit, which is capped at one million in the default verifier. When implementing compaction in eBPF, the verifier explores all possible merge states in the compaction logic, significantly increasing the verification instruction count. However, this limit is intended to bound the complexity and termination time of the verification process itself, rather than to constrain the runtime or the number of executed instructions of an eBPF program. Relaxing the instruction count limit does not weaken the verifier’s core safety checks. The verifier continues to analyze loop constructs for termination and rejects programs with potential infinite loops. Therefore, lifting this limit expands the scope of verification without compromising termination guarantees.

Second, we allow eBPF programs to access a kernel memory region that stores data required for compaction, while strictly restricting accesses to a predefined address range. Since the eBPF stack size is limited, it cannot provide sufficient space to perform merge operations at kilobyte granularity. To prevent arbitrary accesses to kernel memory, we customize the verifier’s \texttt{is\_valid\_access} logic to permit accesses only to the addresses explicitly designated by \sol{}. Because the verifier explores all possible program states, it rejects any eBPF program whose control flow could access outside the allowed kernel memory region. This design eliminates the possibility of unintended kernel memory accesses.

\noindent\textbf{Deterministic IO via SST-Map.}
SST-Map executes user-registered IO operations sequentially and exactly once. In contrast, non-deterministic execution patterns such as data chasing, where the sequence of IO operations depends on runtime data, do not guarantee completion in finite time. By adopting a simple model that performs each IO once in a fixed order, \sol{} only needs to track which IOs have been completed and which remain pending in the SST-Map. IO errors at file descriptors or data offsets are safely handled by the underlying io\_uring subsystem. As a result, \sol{} can execute user logic within the kernel while maintaining kernel stability and retaining precise control over IO execution.

\vspace{-4pt}
\subsection{Detailed Operation of \sol{}}
\label{sec:design_datail}
\noindent\textbf{Start of \texttt{ReadNextKV()}.}
We now detail a single \texttt{ReadNextKV()} invocation. As shown in Figure~\ref{fig:design_solution_detail}(a), the SST-Map is first passed to io\_uring. The SST-Map contains the index, offset, and size of each data block to be compacted for every target SSTable (SST0, SST1, …). Because both eBPF and io\_uring are managed via file descriptors, \sol{} can create an independent instance per compaction thread, with per-thread SST-Maps in separate spaces.

Using the SST-Map, the controller initially reads the first data block (SST0-0, SST1-0, …, SSTn-0) of each target SSTable into kernel memory via io\_uring. All necessary blocks can be requested at once to reduce overhead and increase parallelism. The eBPF program then performs a conventional iterator-style merge-sort over the in-kernel data blocks and writes the results to a kernel-side write buffer.
At this stage, \sol{} can directly access the designated kernel buffer region without relying on helper functions, thereby eliminating helper function call overhead during execution.

\noindent\textbf{Data Block Pointers and End-of-Block Handling.}
Because the SST-Map provides only metadata and the controller does not execute compaction itself, it must track progress. The eBPF program maintains a per-block read pointer and shares it with the controller (Figure~\ref{fig:design_solution_detail}(b)). When the pointer for block SST0-0 reaches its end, the controller consults the SST-Map to locate the next block and issues an io\_uring read. Multiple exhausted blocks can be serviced concurrently by batching these requests.

\noindent\textbf{End of \texttt{ReadNextKV()} and \texttt{WriteKV()}.}
\sol{} iterates by (i) issuing I/O via the controller and io\_uring, and (ii) performing merge-sort and appends via the eBPF program. The loop terminates when (a) the kernel write buffer reaches a predefined threshold or (b) all SSTables are fully merged and no further blocks remain (Figure~\ref{fig:design_solution_detail}(c)). In both cases, the write buffer—resident in user–kernel shared memory—is returned to user space, and \texttt{ReadNextKV()} returns its size. Finally, \texttt{WriteKV()} persists data using the conventional path (Figure~\ref{fig:design_solution_detail}(d)). The user code decides whether to reinvoke \texttt{ReadNextKV()} based on the returned size.

\vspace{-4pt}
\subsection{Merge-sort Algorithm}
\label{sec:design_algo}
\sol{} ports the compaction merge-sort into kernel space via an eBPF program. The algorithm compares keys across multiple data blocks to select the next key–value pair. Users can provide custom comparators or filtering logic and execute them in the kernel via eBPF. While RocksDB typically relies on a min-heap due to its merge-iterator structure, \sol{} is not bound to that choice. Since heap management overhead can be non-trivial for a small number of files, we consider two implementations for minimum selection.

\noindent\textbf{Linear Search.}
Linear search scans all valid data blocks at each iteration and compares their keys. Because it involves few branches, it is efficient when the number of files is small; however, its cost increases proportionally with the number of files and key length. Pseudocode is shown in Algorithm~\ref{alg:next_linear}.

\noindent\textbf{Min-heap Search.}
Min-heap search initializes a min-heap with the current key of each data block at the start of compaction. Each iteration pops the smallest key and pushes the next key from the same block, recording blocks that require additional reads for reinsertion in subsequent calls. A dedicated BPF map preserves the heap across eBPF invocations. With a heap, the per-iteration cost is $O(K\log N)$ for $N$ files and key size $K$, enabling fast selection for large $N$; however, for small $N$, heap maintenance overhead can make it slower than linear search. Pseudocode is given in Algorithm~\ref{alg:next_minheap}.
\section{Evaluation}
\label{sec:eval}

\begingroup
\setlength{\textfloatsep}{0pt}  
\begin{figure*}[t]
  \centering
  \captionsetup[subfigure]{labelformat=empty}
  \subfloat[]{%
    \includegraphics[width=0.6\linewidth]{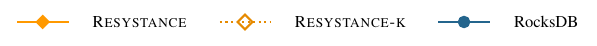}}
  \vspace{-20pt}
  \\[0pt]

  \includegraphics[width=0.27\linewidth]{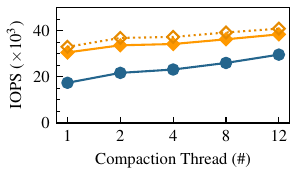}
  \hspace{12pt}
  \includegraphics[width=0.28\linewidth]{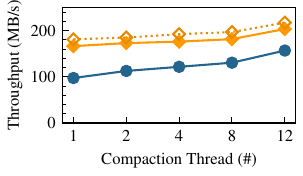}
  \hspace{12pt}
  \includegraphics[width=0.27\linewidth]{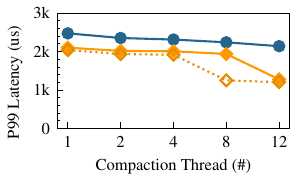}

  \makebox[0.27\linewidth]{{ (a) IO Throughput}}
  \hspace{12pt}
  \makebox[0.28\linewidth]{{ (b) Compaction Throughput}}
  \hspace{12pt}
  \makebox[0.27\linewidth]{{ (c) Latency}}

  \caption{ Performance comparison under the FillRandom workload with varying numbers of compaction threads: overall IOPS, compaction throughput, and latency.}
  \label{fig:eval_fr}
  \vspace{-12pt}
\end{figure*}
\endgroup

\vspace{-4pt}
\subsection{Implementation}
\label{sec:eval_impl}
\sol{} is implemented based on the io\_uring system call in Linux kernel 6.14, with minor kernel modifications to add attachment points for eBPF invocation. Thus, \sol{} operates as an io\_uring instance performing compaction by using a dedicated flag, without introducing any new system calls. The eBPF program employs the \texttt{bpf\_loop} helper function to enable the verifier to safely validate moderately complex compaction logic. In this study, the threshold number of input files for selecting the search algorithm is empirically set to six: when the number of compaction files is less than or equal to six, the Linear Search algorithm is used; otherwise, the Min-heap Search is applied. Detailed evaluation of the algorithm is discussed in evaluation Section~\ref{sec:eval_features}.

RocksDB~\cite{rocksdb} version 10.1.3 is used as the baseline. We modified the \texttt{compaction\_job.cc} source file, where the compaction job is implemented, to invoke \sol{} by calling the \texttt{io\_uring\_enter} system call with a special flag instead of using \texttt{c\_iter()}. Within the same function, the \texttt{io\_uring\_setup} system call initializes the io\_uring instance, and the \texttt{mmap} system call maps shared user–kernel memory for compaction data exchange. Initialization occurs once per compaction thread to maximize reusability and minimize invocation overhead. Each compaction thread maintains its own independent io\_uring instance (i.e., file descriptor) to perform parallel and isolated compaction jobs.

The source code for \sol{} is publicly available at \url{https://github.com/DISCOS-LAB/Resystance}.

\SetKwInOut{Input}{Input}
\SetKwInOut{Output}{Output}
\begin{algorithm}[h]
\scriptsize
\setlength{\tabcolsep}{-20pt}
\caption{\\ \small \texttt{NextLinear}(\textit{data\_block}, \textit{ptr}, \textit{write\_buffer})}
\label{alg:next_linear}
\Input{array \textit{data\_block}[1..n]; array \textit{ptr}[1..n]; \textit{write\_buffer}}
\Output{\textit{write\_buffer\_size}}
\DontPrintSemicolon
\SetKwFunction{FMain}{NextLinear}
\SetKwFunction{KeyAt}{KeyAt}
\SetKwFunction{KVAt}{KVAt}
\SetKwFunction{KVSize}{KVPairSizeAt}
\SetKwFunction{Append}{Append}
\SetKwProg{Fn}{Function}{:}{}

\Fn{\FMain{$\textit{data\_block},\, \textit{ptr},\, \textit{write\_buffer}$}}{
    $ \textit{Valid} \gets$ $\{\, i \in [1..n] \mid \textit{data\_block}[i] \text{ is valid} \,\}$ \;
    \While{$\forall i \in \textit{Valid},~ \textit{ptr}[i] < \textit{block\_size}[i]$}{
        $\textit{idx} \gets -1$; $\textit{best\_key} \gets NULL$ \;
        \ForEach{$i \in \textit{Valid}$}{
            $\textit{key} \gets \KeyAt{data\_block[i],\, ptr[i]}$ \;
            \If{$\textit{idx} = -1$ OR $\texttt{memcmp}(\textit{key},\, \textit{best\_key}) < 0$}{
                $\textit{idx} \gets i$; $\textit{best\_key} \gets \textit{key}$ \;
            }
        }
        $\textit{adv} \gets \KVSize{data\_block[idx],\, ptr[idx]}$ \;
        $\textit{kv} \gets \KVAt{data\_block[idx],\, ptr[idx]}$ \;
        $\textit{ptr}[\textit{idx}] \gets \textit{ptr}[\textit{idx}] + \textit{adv}$ \;
        \Append{\textit{write\_buffer},\, \textit{kv}} \;
    }
    \Return $\lvert \textit{write\_buffer} \rvert$ \;
}
\end{algorithm}

\SetKwInOut{Input}{Input}
\SetKwInOut{Output}{Output}
\begin{algorithm}[h]
\scriptsize
\caption{\\ \small \texttt{NextMinHeap}(\textit{data\_block}, \textit{ptr}, \textit{write\_buffer}, \textit{H})}
\label{alg:next_minheap}
\Input{array $\textit{data\_block}[1..n]$; array $\textit{ptr}[1..n]$; \textit{write\_buffer}; minheap \textit{H}}
\Output{\textit{write\_buffer\_size}}
\DontPrintSemicolon
\SetKwFunction{FMain}{NextMinHeap}
\SetKwFunction{KeyAt}{KeyAt}
\SetKwFunction{KVAt}{KVAt}
\SetKwFunction{KVSize}{KVPairSizeAt}
\SetKwFunction{Append}{Append}
\SetKwFunction{Push}{Push}
\SetKwFunction{Pop}{Pop}
\SetKwFunction{Empty}{IsEmpty}
\SetKwProg{Fn}{Function}{:}{}

\Fn{\FMain{$\textit{data\_block},\, \textit{ptr},\, \textit{write\_buffer},\, \textit{H}$}}{
    $\textit{Valid} \gets \{\, i \in [1..n] \mid \textit{data\_block}[i] \text{ is valid} \,\}$ \;

    \If{first call}{
        \ForEach{$i \in \textit{Valid}$}{
            $\textit{key} \gets \KeyAt{data\_block[i],\, ptr[i]}$ \;
            \Push{$\textit{H},\,(\textit{key},\, i)$} \;
        }
    }
    \If{$\texttt{prev\_pop\_block} \ne \texttt{NULL}$}{
        $j \gets \texttt{prev\_pop\_block}$ \;
        $\texttt{prev\_pop\_block} \gets \texttt{NULL}$ \;
        $\textit{key} \gets \KeyAt{\textit{data\_block}[j],\, \textit{ptr}[j]}$ \;
        \Push{$\textit{H},\,(\textit{key},\, j)$} \;
    }

    \While{\texttt{true}}{
        $(\textit{key},\, \textit{idx}) \gets \Pop{\textit{H}}$ \;
        $\textit{kv} \gets \KVAt{\textit{data\_block}[idx],\, \textit{ptr}[idx]}$ \;
        $\textit{adv} \gets \KVSize{\textit{data\_block}[\textit{idx}],\, \textit{ptr}[\textit{idx}]}$ \;
        $\textit{ptr}[\textit{idx}] \gets \textit{ptr}[\textit{idx}] + \textit{adv}$ \;
        \Append{\textit{write\_buffer},\, \textit{kv}} \;
        \If{$\textit{ptr}[\textit{idx}] < \textit{block\_size}[\textit{idx}]$}{
            $\textit{key'} \gets \KeyAt{\textit{data\_block}[idx],\, \textit{ptr}[idx]}$ \;
            \Push{$\textit{H},\,(\textit{key'},\, \textit{idx})$} \;
        }
        \Else{
            $\texttt{prev\_pop\_block} \gets \textit{idx}$ \;
            \textbf{break} \;
        }
    }
    \Return $\lvert \textit{write\_buffer} \rvert$ \;
}
\end{algorithm}

\vspace{-4pt}
\subsection{Experimental Setup}
\label{sec:expr_setup}

\noindent\textbf{Platform.}
All experiments were conducted on a server with an AMD Ryzen 9 7950X @ 4.5~GHz CPU (SMT and C-states disabled), 32~GB DDR5 DRAM, and a Samsung 990 PRO 1~TB NVMe SSD. The system runs Ubuntu~24.04.3 with the Linux~6.14 kernel and an ext4 file system.

\noindent\textbf{Comparison.}
We compare \sol{} against the baseline RocksDB, as \sol{} optimizes only the software I/O stack without altering the LSM-tree structure or compaction algorithm. This makes RocksDB a fair and meaningful baseline for evaluating \sol{}’s effectiveness.  
Additionally, to evaluate eBPF execution overhead and in-kernel efficiency, we include \solk{}, a variant of \sol{} that integrates the eBPF code entirely into the kernel and executes compaction directly within io\_uring.  
The evaluation targets are as follows:

\squishlist
\item \textbf{RocksDB:} The baseline configuration using conventional system calls through the kernel path during compaction.
\item \textbf{\sol{}:} Our proposed system reduces software stack overhead during compaction using {io\_uring} and eBPF.
\item \textbf{\solk{}:} A kernel-integrated variant of \sol{}, where the eBPF logic is embedded directly into the kernel, and compaction is executed entirely within {io\_uring}.
\squishend

\noindent\textbf{RocksDB Parameter Configurations.}
To isolate compaction performance effects, we disable the Write-Ahead Log, Block Cache, and compression. Direct I/O is used for both compaction and I/O requests. The compaction policy follows the default Leveled strategy, with L0-to-L0 intra compaction disabled to eliminate write stall mitigation effects. Two memtables of 64~MB each and one flush thread are used.

\noindent\textbf{Benchmarks and Workloads.}
We evaluate \sol{} using three representative benchmarks:  
First, db\_bench~\cite{dbbench} is used for synthetic workload analysis.  
Second, we employ the YCSB benchmark~\cite{cooper2010benchmarking} to evaluate real-world read–write access patterns in key-value store environments.  
In db\_bench and YCSB, all workloads use 16~B keys and 1~KB values.
Furthermore, we evaluate the effectiveness of \sol{} in a RocksDB-based relational database management system (RDBMS) using Sysbench's OLTP workload.
The workloads are defined as follows:

\squishlist
\item \textbf{FillRandom:} A 100\% write workload in db\_bench.
\item \textbf{ReadRandomWriteRandom:} A mixed read–write workload with variable read/write ratios in db\_bench, executed after data loading via FillRandom.
\item \textbf{ReadWhileWriting:} 100\% read thread and 100\% write thread each run concurrently in db\_bench and are executed after FillRandom.
\item \textbf{YCSB:} Standard YCSB workloads as described in Table~\ref{tbl:workload_ycsb}.
\item \textbf{OLTP:} OLTP workloads are evaluated using Sysbench~\cite{sysbench} on a RocksDB-based RDBMS. Detailed workload descriptions are provided in Table~\ref{tbl:workload_oltp}.
\squishend

In FillRandom, ReadRandomWriteRandom, and YCSB, each I/O thread performs operations on 10M entries. In ReadWhileWriting and OLTP workloads, operations are performed for 1200 seconds.

\begingroup
\begin{table}[b]
\renewcommand{\arraystretch}{1}
    \centering 

    \caption{Description of YCSB Workload Characteristics.}
    \vspace{-4pt}

    \resizebox{0.9\columnwidth}{!}{
    \tiny{
\begin{tabular}{cccc}
\toprule
\textbf{Type} & \textbf{Description} & \textbf{Distribution}  \\ \midrule
Load & 100\% Writes & Zipfian  \\ 
A & 50\% Updates, 50\% Reads & Zipfian  \\ 
B & 95\% Reads, 5\% Updates   & Zipfian  \\ 
C & 100\% Reads & Zipfian  \\ 
D & 95\% Reads, 5\% Inserts & Latest  \\ 
E & 95\% Seeks, 5\% Inserts; Scan length:128 & Zipfian  \\ 
F & 50\% Read-Mondify-Write, 50\% Reads & Zipfian  \\ \bottomrule
\end{tabular}
    }
    }
    \label{tbl:workload_ycsb}
\end{table}
\endgroup

\begingroup
\setlength{\textfloatsep}{0pt}  
\begin{figure*}[t]
  \centering
  \captionsetup[subfigure]{labelformat=empty}
  \subfloat[]{
    \includegraphics[width=0.6\linewidth]{plots/eval_fr_legend.pdf}}
  \vspace{-20pt}
  \\[0pt]

  \hfill
  \parbox[b]{0.03\linewidth}{
    \centering
    \rotatebox{90}{\textbf{$\qquad$IO Throughput}}
  }\hfill
  \includegraphics[width=0.23\linewidth]{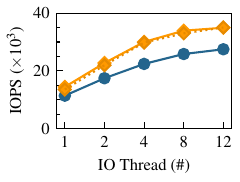}
  \hfill
  \includegraphics[width=0.23\linewidth]{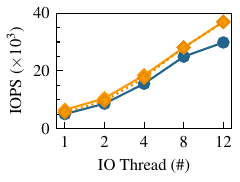}
  \hfill
  \includegraphics[width=0.23\linewidth]{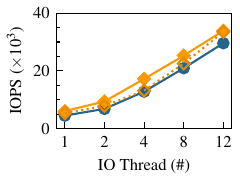}
  \hfill
  \includegraphics[width=0.23\linewidth]{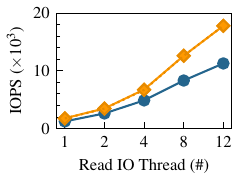}

  \hfill
  \parbox[b]{0.03\linewidth}{
    \centering
    \rotatebox{90}{\textbf{$\qquad$Read Latency}}
  }\hfill
  \includegraphics[width=0.23\linewidth]{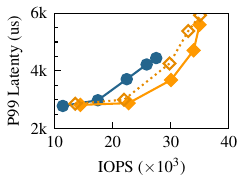}
  \hfill
  \includegraphics[width=0.23\linewidth]{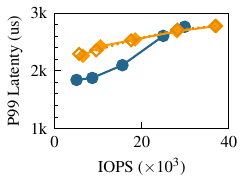}
  \hfill
  \includegraphics[width=0.23\linewidth]{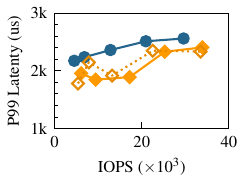}
  \hfill
  \includegraphics[width=0.23\linewidth]{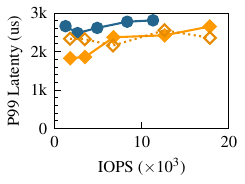}
    
  \hfill
  \parbox[b]{0.03\linewidth}{
    \centering
    \rotatebox{90}{\textbf{$\qquad$Write Latency}}
  }\hfill
  \includegraphics[width=0.23\linewidth]{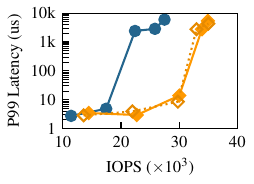}
  \hfill
  \includegraphics[width=0.23\linewidth]{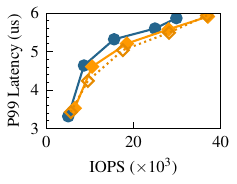}
  \hfill
  \includegraphics[width=0.23\linewidth]{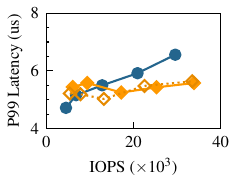}
  \hfill
  \includegraphics[width=0.23\linewidth]{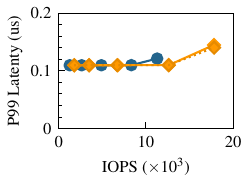}

  \hfill
  \makebox[0.24\linewidth]{{(a) R10/W90}}
  \hfill
  \makebox[0.24\linewidth]{{(b) R50/W50}}
  \hfill
  \makebox[0.24\linewidth]{{(c) R90/W10}}
  \hfill
  \makebox[0.24\linewidth]{{(d) ReadWhileWriting}}

  \caption{Performance comparison varying the number of IO threads with 4 compaction threads. (a)--(c) Variations in read/write ratio under the ReadRandomWriteRandom workload, (d) Increase in the number of Read IO threads under the ReadWhileWriting workload.
  }
  \label{fig:eval_rrwr}
  \vspace{-12pt}
\end{figure*}
\endgroup

\vspace{-4pt}
\subsection{Overall Performance}
\vspace{-4pt}
\subsubsection{Write-Intensive Workload}
Figure~\ref{fig:eval_fr} shows the results of the FillRandom workload with varying numbers of compaction threads.  
As shown in Figure~\ref{fig:eval_fr}(a), both \sol{} and \solk{} achieve higher IOPS than RocksDB, with \sol{} delivering up to 75\% higher throughput compared to RocksDB using a single compaction thread.
\solk{} slightly outperforms \sol{} due to the inherent invocation and execution overhead of the eBPF mechanism. 
Figure~\ref{fig:eval_fr}(b) shows compaction throughput, which closely mirrors the IOPS trend—confirming that throughput improvement stems from faster compaction processing. On average, \sol{} reduces system call invocations by over 99\% and shortens compaction execution time by about 50\%.
Figure~\ref{fig:eval_fr}(c) illustrates p99 write latency, where both \sol{} and \solk{} outperform RocksDB. With 12 compaction threads, \sol{} achieves up to 40\% lower p99 latency than RocksDB. The larger gap between \sol{} and \solk{} at eight threads occurs because even small compaction differences can cause significant latency variations under write stall conditions.

\subsubsection{Read-Write Mixed Workload}
Figure~\ref{fig:eval_rrwr}(a)--(c) shows results for ReadRandomWriteRandom workloads with four compaction threads while varying the number of I/O threads. The first row shows IOPS, and the second and third rows show p99 read and write latency, respectively.

For R10/W90 (Figure~\ref{fig:eval_rrwr}(a)), throughput follows a similar trend to FillRandom, but the performance gap between \sol{} and \solk{} nearly disappears, indicating negligible eBPF overhead when compaction frequency decreases. For p99 read latency, both \sol{} and \solk{} achieve lower latency than RocksDB at comparable throughput. As throughput saturates, latency slightly increases but remains below that of RocksDB.  
The p99 write latency of \sol{} and \solk{} is significantly lower (especially when normalized by throughput) than RocksDB, due to the effective mitigation of write stalls.

Figure~\ref{fig:eval_rrwr}(b) shows R50/W50 results. As the write ratio drops, the throughput gap between \sol{} and RocksDB narrows. At low throughput, \sol{} and \solk{} show slightly higher p99 read latency than RocksDB as faster compaction increases disk access concurrency. However, as throughput increases, faster compaction stabilizes read performance. p99 write latency differences are minor, as write stalls play a smaller role in mixed workloads.

For R90/W10 (Figure~\ref{fig:eval_rrwr}(c)), throughput trends remain similar to R50/W50, with \sol{} maintaining higher throughput and lower p99 latency in both reads and writes, demonstrating the benefit of efficient compaction even in read-intensive workloads.

\begingroup
\setlength{\textfloatsep}{0pt}  
\begin{figure*}[t]
  \centering
  \captionsetup[subfigure]{labelformat=empty}
  \subfloat[]{
    \includegraphics[width=0.6\linewidth]{plots/eval_fr_legend.pdf}}
  \vspace{-20pt}
  \\[0pt]

  \includegraphics[width=0.24\linewidth]{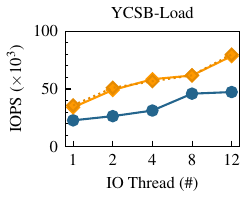}
  \includegraphics[width=0.24\linewidth]{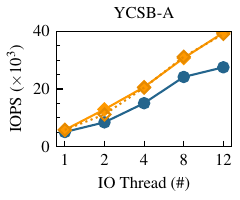}
  \includegraphics[width=0.24\linewidth]{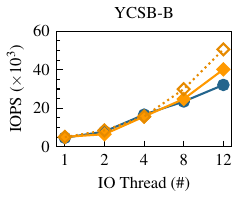}
  \hfill
  \\
  \includegraphics[width=0.24\linewidth]{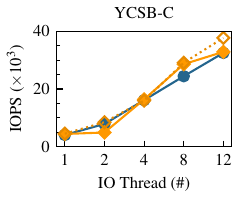}
  \hfill
  \includegraphics[width=0.24\linewidth]{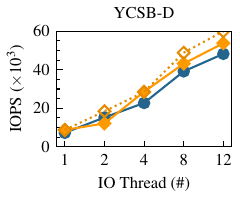}
  \hfill
  \includegraphics[width=0.24\linewidth]{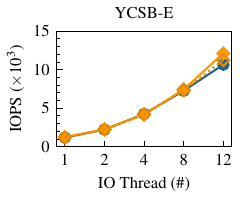}
  \hfill
  \includegraphics[width=0.24\linewidth]{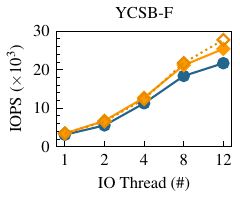}
  \vspace{-6pt}
  \caption{Comparison of throughput across YCSB benchmark workloads.}
  \label{fig:eval_ycsb}
\vspace{-12pt}
\end{figure*}
\endgroup

\begingroup
\setlength{\textfloatsep}{0pt}  
\begin{figure*}[t]
  \centering
  \captionsetup[subfigure]{labelformat=empty}
  \subfloat[]{%
    \includegraphics[width=0.6\linewidth]{plots/eval_fr_legend.pdf}}
  \vspace{-20pt}
  \\[0pt]

  \includegraphics[width=0.24\linewidth]{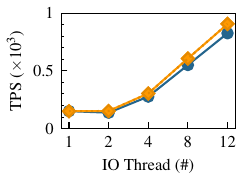}
  \hfill
  \includegraphics[width=0.24\linewidth]{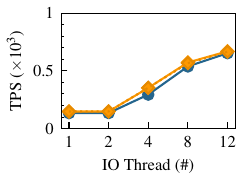}
  \hfill
  \includegraphics[width=0.24\linewidth]{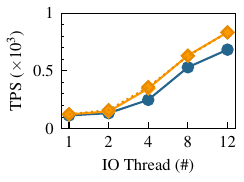}
  \hfill
  \includegraphics[width=0.24\linewidth]{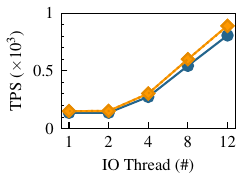}

    \hspace{-3pt}
  \makebox[0.24\linewidth]{{(a) oltp\_insert}}
  \hfill
  \makebox[0.24\linewidth]{{(b) oltp\_write\_only}}
  \hfill
  \makebox[0.24\linewidth]{{(c) oltp\_read\_write}}
  \hfill
  \makebox[0.24\linewidth]{{(d) oltp\_update\_non\_index}}
  
  \caption{Sysbench evaluation of MariaDB with MyRocks.}
  \label{fig:eval_rdbms}
  \vspace{-12pt}
\end{figure*}
\endgroup

\subsubsection{Read-Intensive Workloads under Concurrent Writes}
\label{sec:eval_rww}

Figure~\ref{fig:eval_rrwr}(d) shows the performance of ReadWhileWriting with four compaction threads as the number of read IO threads increases. As the number of read IO threads grows, the throughput improvement of \sol{} becomes more pronounced, achieving up to 57\% higher throughput. This result indicates that improved compaction throughput positively affects read-intensive workloads. Moreover, even as throughput increases, \sol{} maintains more stable p99 read latency compared to RocksDB.

\subsubsection{YCSB Workloads}
Figure~\ref{fig:eval_ycsb} presents the throughput comparison under YCSB workloads using four compaction threads. As summarized in Table~\ref{tbl:workload_ycsb}, \sol{} achieves up to 85\% higher throughput than RocksDB in YCSB-Load, surpassing even the FillRandom workload. This improvement arises from the Zipfian access pattern: higher key-drop ratios amplify \sol{}’s advantage of in-kernel processing.  
In YCSB-A, which has a high write ratio, \sol{} achieves up to 53\% higher throughput than RocksDB. In read-dominant and range-scan workloads, \sol{} consistently outperforms RocksDB by a smaller margin, as reduced compaction frequency limits the effect of compaction acceleration. Nevertheless, as the number of I/O threads increases, the throughput gap widens—underscoring that \sol{}’s compaction efficiency scales effectively with I/O parallelism.

\begin{table}[b]
    \centering 
    \caption{Description of OLTP workload characteristics.}
    \vspace{-4pt}
    \resizebox{\columnwidth}{!}{
    \tiny{
\begin{tabular}{c|c|c|c}
\toprule
\textbf{Workload Type} &
\begin{tabular}[c]{@{}c@{}}
\textbf{Operations per Transaction} \\
\textbf{(Execution Order)}
\end{tabular} &
\begin{tabular}[c]{@{}c@{}}
\textbf{Key (Bytes)} \\
\textbf{range / avg.}
\end{tabular} &
\begin{tabular}[c]{@{}c@{}}
\textbf{Value (Bytes)} \\
\textbf{range / avg.}
\end{tabular}
\\ \midrule

oltp\_insert &
1 \texttt{INSERT} &
\multirow{4}{*}{4--42 / 8} &
\multirow{4}{*}{2--724 / 722}
\\

oltp\_write\_only &
2 \texttt{UPDATE}, 1 \texttt{DELETE}, 1 \texttt{INSERT} &
&
\\

oltp\_read\_write &
\begin{tabular}[c]{@{}l@{}}
14 \texttt{SELECT}, 2 \texttt{UPDATE}, \\
1 \texttt{INSERT}, 1 \texttt{DELETE}
\end{tabular} &
&
\\

oltp\_update\_non\_index &
1 \texttt{UPDATE} &
&
\\

\bottomrule
\end{tabular}
    }
    }
    \label{tbl:workload_oltp}
\end{table}

\subsubsection{OLTP Workloads on RDBMS}
\label{sec:eval_rdbms}
Figure~\ref{fig:eval_rdbms} presents an evaluation of OLTP workloads on MariaDB, an RDBMS that uses RocksDB as its storage engine (MyRocks~\cite{matsunobu2020myrocks}), using Sysbench. The transaction layer merely interfaces with RocksDB, while compaction is executed entirely within the RocksDB storage engine; therefore, once RESYSTANCE is integrated into RocksDB, neither RocksDB’s upper layers nor MariaDB’s transaction layer require further modification. Unlike the previous experiments described in Section~\ref{sec:expr_setup}, this evaluation enables Write-Ahead Logging (WAL) and L0-to-L0 intra-compaction to assess the practical effectiveness of \sol{} in an RDBMS setting (with four compaction threads). In addition, as shown in Table~\ref{tbl:workload_oltp}, the OLTP workloads involve diverse key distributions and value sizes. Except for oltp\_write\_only (Figure~\ref{fig:eval_rdbms}(c)), the throughput gap between \sol{} and RocksDB widens as the number of IO threads increases, demonstrating the effectiveness of \sol{}. 
In particular, \sol{} achieved 22\% higher throughput than RocksDB in oltp\_read\_write.
The {oltp\_write\_only} workload exhibits read-modify-write behavior, making its performance primarily dependent on transaction processing overhead; consequently, the benefits of improved compaction are relatively limited.

\subsection{Analysis of \sol{} Features}
\label{sec:eval_features}
\vspace{-4pt}

\subsubsection{Comparison of Merge-sort Algorithms}
Figure~\ref{fig:eval_mergesort} compares compaction execution time across different merge-sort algorithms in \sol{} as a function of the number of SST files. We evaluate two representative value sizes: 1KB, which is commonly used in benchmarks, and 128B, which is frequently observed in real-world RocksDB deployments~\cite{cao2020characterizing}. For both value sizes, the linear merge-sort slightly outperforms the min-heap approach when merging up to 6-8 SSTs, after which the min-heap approach becomes faster. These results justify the merge-sort selection strategy adopted by \sol{} in Section~\ref{sec:eval_impl}.
However, these marginal performance differences do not have a dominant impact on overall system throughput, as compaction involves diverse SST file distributions and significant time is spent in stages other than merging. Nevertheless, by evaluating two simple algorithms, we demonstrate that \sol{} leaves room for further optimization due to its underlying architecture.

\subsubsection{eBPF Program Verification Overhead}
\label{sec:eval_ebpf}
We analyzed the verification overhead of the eBPF program to show that the verification process for loading the kernel of the user-defined merge logic performed correctly. 
Figure~\ref{fig:eval_verification} shows the verification overhead incurred when loading the linear merge-sort eBPF program into the kernel as a function of the maximum number of allowed SST files. Figure~\ref{fig:eval_verification}(a) reports verification time, which increases exponentially with the maximum SST file count due to exponential growth of the verifier’s state space at branch points. Figure~\ref{fig:eval_verification}(b) presents the number of verified instructions on a logarithmic scale and shows a similar exponential trend, exceeding the default verification limit of 1M when the number of SST files reaches 24.
In contrast, the min-heap–based implementation incurs a negligible verification cost, with verification time bounded by 0.05s and instruction counts ranging from 20K to 100K, well below the verification instruction limit. This is because the min-heap algorithm involves loops with a small and statically bounded number of iterations, resulting in a much smaller verification state space. Both implementations require stack sizes well below the eBPF limit (64B for linear and 128B for min-heap), as merge-related data are stored and accessed through kernel memory rather than the eBPF stack.

\subsubsection{Impact of Key, Value, and Compaction Input Sizes}
\label{sec:eval_impact_compaction}
Figure~\ref{fig:eval_effect}(a)--(c) show the effectiveness of \sol{} under varying key, value sizes and compaction input sizes, respectively. The $y$-axis shows compaction time normalized to RocksDB.
Figure~\ref{fig:eval_effect}(a) and (b) show the results when the key size varies (with the value fixed at 1KB) and when the value size varies (with the key fixed at 16B), respectively. \sol{} reduces compaction time by approximately 50\% regardless of key or value size. This is because key and value sizes only affect the number of key–value pairs within a data block, but do not influence the I/O path of \sol{}.
Figure~\ref{fig:eval_effect}(c) presents compaction time as a function of compaction input size for 1KB values, where smaller compaction inputs yield larger performance gains. This trend arises because, as compaction overhead decreases, software stack overhead becomes more pronounced, thereby amplifying the relative effectiveness of \sol{}.

\begingroup
\setlength{\textfloatsep}{-10pt} 
\begin{figure}[t]
  \centering

  \captionsetup[subfigure]{labelformat=empty}
  \subfloat[]{%
    \includegraphics[width=1\linewidth]{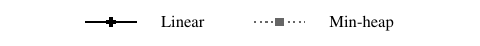}}
  \vspace{-20pt}
  \\[0pt]

  \hspace{-3pt}
  \includegraphics[width=0.47\linewidth]{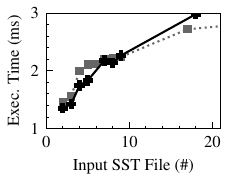}
  \includegraphics[width=0.5\linewidth]{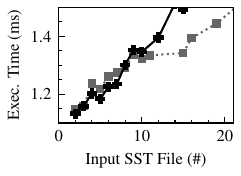}
  \hspace{-3pt}
  
  \makebox[0.47\linewidth]{{(a) 128B}}
  \makebox[0.5\linewidth]{{(b) 1KB}}

  \caption{
  Comparison of compaction execution times by merge-sort algorithm for different value sizes.
  }
  \vspace{-8pt}
  \label{fig:eval_mergesort}
\end{figure}
\endgroup

\begingroup
\setlength{\textfloatsep}{0pt}  
\begin{figure}[!t]
  \centering

  \captionsetup[subfigure]{labelformat=empty}

  \hspace{-3pt}
  \includegraphics[width=0.47\linewidth]{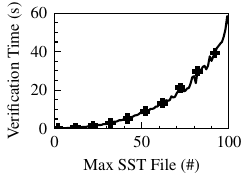}
  \includegraphics[width=0.47\linewidth]{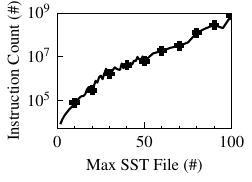}

  \hspace{-3pt}
  \makebox[0.47\linewidth]{{(a) Verification Time}}
  \makebox[0.5\linewidth]{{(b) Verification Instruction Count}}

  \caption{
  Verification overhead of eBPF programs with varying limits on the maximum number of SSTs allowed. (a) Verification time, (b) Verification instruction count (log scale).
  }

  \label{fig:eval_verification}
  \vspace{-6pt}
\end{figure}
\endgroup

\begingroup
\setlength{\textfloatsep}{0pt}  
\begin{figure}[t]
  \centering

  \captionsetup[subfigure]{labelformat=empty}
  \subfloat[]{
    \includegraphics[width=1\linewidth]{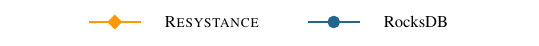}}
  \vspace{-20pt}
  \\[0pt]

  \hspace{-6pt}
  \includegraphics[width=0.33\linewidth]{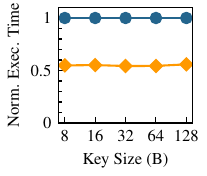}
  \hspace{-6pt}
  \includegraphics[width=0.33\linewidth]{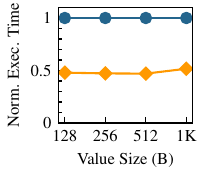}
  \includegraphics[width=0.33\linewidth]{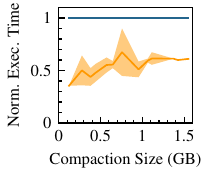}

  \hspace{-6pt}
  \makebox[0.33\linewidth]{{(a) Key Size}}
  \hspace{-6pt}
  \makebox[0.33\linewidth]{{(b) Value Size}}
  \makebox[0.33\linewidth]{{(c) Input Size}}

  \caption{
  Compaction effectiveness of \sol{} for various key, value, and input sizes. The $y$-axis represents the normalized compaction processing time in RocksDB. (a) Comparison by key size, (b) Comparison by value size, (c) Comparison by compaction input size. 
  }
  \label{fig:eval_effect}
\vspace{-12pt}
\end{figure}
\endgroup

\begingroup
\setlength{\textfloatsep}{0pt}  
\begin{figure}[t]
  \centering

  \captionsetup[subfigure]{labelformat=empty}

  \hspace{-3pt}
  \includegraphics[width=0.45\linewidth]{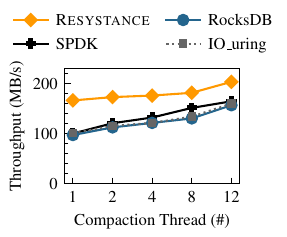}
  \includegraphics[width=0.47\linewidth]{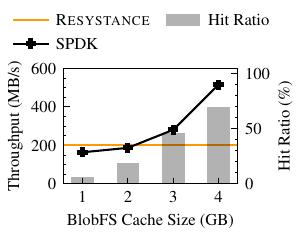}

  \hspace{-3pt}
  \makebox[0.45\linewidth]{{(a) SPDK and io\_uring}}
  \makebox[0.47\linewidth]{{(b) BlobFS Cache}}

  \caption{
  Comparison of compaction throughput with RocksDB, SPDK, and io\_uring. (a) Comparison of \sol{}, SPDK, and io\_uring with varying numbers of compaction threads. (b) Comparison of SPDK with BlobFS with varying cache sizes.
  }
  \vspace{-2pt}
  \label{fig:eval_spdk}
\end{figure}
\endgroup

\subsubsection{Effectiveness of Kernel Bypass and Asynchronous I/O}
\label{sec:eval_spdk}
To evaluate the impact of kernel bypass and asynchronous I/O on compaction, we compare compaction performance across three configurations: vanilla RocksDB, RocksDB with io\_uring, and SPDK-based RocksDB. Figure~\ref{fig:eval_spdk}(a) presents compaction throughput as a function of the number of compaction threads. SPDK achieves kernel bypass using BlobFS~\cite{yang2017spdk}, which maintains a private user-space cache instead of relying on the kernel page cache. This cache cannot be disabled; therefore, we configure the minimum experimentally determined cache size of 1GB. Due to the limited performance of BlobFS in this configuration, the compaction improvement achieved by SPDK is modest. While io\_uring enables compaction acceleration through asynchronous I/O, the inherent characteristics of compaction limit the performance benefits of async I/O, resulting in negligible compaction improvement.
Figure~\ref{fig:eval_spdk}(b) compares throughput and BlobFS cache hit ratio under 12 compaction threads as the BlobFS cache size varies. With a 1GB cache, the cache hit ratio remains very low, indicating that most I/O operations are performed via direct I/O. As the cache size increases, SPDK throughput improves significantly, accompanied by a corresponding increase in cache hit ratio. This result indicates that the observed performance gains primarily stem from increased memory accesses rather than reduced I/O overhead through kernel bypass.
\vspace{-4pt}
\section{Related Work }
\label{sec:related}

\subsection{Improving Compaction in LSM-trees}
\noindent\textbf{Optimizing Compaction Policies.}
Dostoevsky~\cite{Dayan2018Dostoevsky} and Spooky~\cite{Dayan2022Spooky} minimize unnecessary compactions by selecting hybrid merge policies based on cost-model analysis. Studies that aim to reduce compaction volume~\cite{Wu2015Lsmtrie, yao2020matrixkv, Wang2022Blockgrained, Sun2023Finegrained, ding2022trianglekv} refine merge granularity to reduce the overhead. Pipelined Compaction~\cite{Zhang2014Pipelined}, LDC~\cite{Chai2020Adaptive}, bLSM~\cite{Sears2012blsm}, LSbM-tree~\cite{Teng2018Lsbmtree}, and DownForce~\cite{byun2025revisiting} improve compaction methods and scheduling algorithms. dCompaction~\cite{Pan2017dCompaciton} leverages metadata to reduce data I/O during compaction.

\noindent\textbf{Dynamic Resource Management.}
SILK~\cite{Balmau2019SILK}, ADOC~\cite{yu2023adoc}, and SuccinctKV~\cite{Zhang2024Succintkv} dynamically tune I/O pressure and background threads to balance data flow against compute resources. DLC~\cite{Jin2021DLC} delays L0 compaction with its scheduler and spreads I/O consumption via bursty compaction.

\noindent\textbf{Redesigning the LSM Structure.}
WipDB~\cite{Zhao2021Wipdb} and PebblesDB~\cite{Raju2017PebblesDB} vertically partition LSM-trees into smaller components to alleviate internal operation costs. LSM-trie~\cite{Wu2015Lsmtrie} uses hash keys to form unique key ranges, eliminating key-range overlap.

\noindent\textbf{Hardware Acceleration and Offloading.}
FPGA acceleration~\cite{zhang2020fpgacompaction, Sun2020FPGA}, GPU pipelining~\cite{Zhou2024GPUCompaction, Sun2024glsm, Xu2020LUDA}, near-data processing~\cite{Sun2019NDP, Sun2024NDP}, DPU offloading~\cite{Pan2017dCompaction}, KV-SSD collaborating~\cite{kim2025kvaccel}, and cloud/edge offloading~\cite{Li2022FaaS, Yu2024CaaS,kim2024coordinating,kim2025eco} shift computation to devices or external environments to reduce host load.

\noindent\textbf{Key-Value Separation.}
WiscKey~\cite{Lu2016WiscKey}, HashKV~\cite{Helen2018HashKV}, and DiffKV~\cite{Li2021DiffKV} separate keys and values. Although not focused on optimizing compaction itself, these designs reduce the amount of data processed by compaction and thus alleviate compaction overhead.

\vspace{-6pt}
\subsection{In-Kernel I/O Path Optimization using eBPF}
\noindent\textbf{Kernel Offloading within the I/O path.}
Recent research uses eBPF to directly optimize the I/O path by eliminating system call and data-copy overheads at the user–kernel boundary. XRP~\cite{zhong2022xrp} inserts eBPF hooks at the NVMe driver level to execute user-defined storage functions in-kernel, thereby bypassing the conventional storage stack. Userspace Bypass~\cite{Zhou2023Userspacebypass} dynamically identifies contiguous syscall regions and safely translates the corresponding user instructions to run inside the kernel, allowing continued execution without crossing syscall boundaries. This approach removes syscall overhead for I/O-intensive workloads without modifying existing applications.

\noindent\textbf{Extending In-Kernel Data Paths with eBPF.}
eBPF has been extended beyond the I/O stack to memory and caching layers. cach\_ext~\cite{Zhssman2025cacheext} allows applications to define page-cache replacement policies as eBPF programs executed in the kernel. BPF-DB~\cite{butrovich2025bpf} embeds a transactional DBMS in the eBPF to provide kernel-level data management consistency and ensure durability and concurrency for eBPF-based applications. eBPF-based page-fault handling~\cite{Zussman2024Custom} further eliminates transfer costs between kernel and user space by handling page faults in eBPF, overcoming scalability limits of \texttt{userfaultfd()}.

\vspace{-6pt}
\subsection{Asynchronous I/O and System Call Reduction}
\label{sec:related_kvell}
KVell~\cite{lepers2019kvell} reduces system call overhead and leverages Linux asynchronous I/O to batch requests and better utilize NVMe parallelism, achieving high throughput even under random-access workloads. However, it attains these gains by relaxing data ordering and redesigning the key-value store architecture. In contrast, our work preserves the existing LSM-tree engine and compaction framework, and instead removes compaction system call overhead using eBPF and io\_uring.

\vspace{-4pt}
\section{Discussion}
\label{sec:discuss}
\noindent\textbf{Why Only the Merge Phase is Designed in eBPF.}
As discussed earlier, applying eBPF programs to compaction requires design choices regarding whether the read and write logic should be included in the eBPF JIT compilation along with the merge logic. In this work, we apply eBPF exclusively to the merge phase.
For writes, during compaction, output data are accumulated in memory and written to disk in large batches. Because such writes are issued infrequently and at large granularity, I/O latency dominates over system call overhead. As a result, applying \sol{} to the write path is unlikely to yield meaningful performance improvements.
Applying eBPF to reads is infeasible due to current limitations of eBPF. First, eBPF does not provide helper functions for initiating disk I/O. Introducing such support would require substantial kernel modifications and potentially weaken verifier guarantees. Second, invoking multiple eBPF programs requires tail calls, significantly increasing programming complexity. Both issues negatively affect system stability and maintainability.
Nevertheless, if these limitations are addressed in the future, it may be possible to eliminate redundant overhead between read and merge phases, enabling further performance improvements.

\noindent\textbf{Maintainability of \sol{}.}
To implement \sol{}, we introduce a dedicated eBPF hook within io\_uring. This design ensures that verifier modifications required by \sol{} are isolated and do not affect other eBPF programs. Moreover, io\_uring largely preserves its existing architecture; \sol{} extends functionality without modifying I/O execution after submission queue entry registration, thereby minimizing interference with other kernel components.
To improve robustness against potential io\_uring API changes, \sol{} interacts only with higher-level abstractions within io\_uring, such as submission queue registration and completion queue handling, rather than relying on internal implementation details.
Consequently, this design maintains compatibility across kernel versions and simplifies long-term maintenance in the presence of kernel evolution.

\noindent\textbf{Adaptability of \sol{}.}
Compaction strategies in LSM-trees are broadly categorized into leveled, tiered, and hybrid approaches, differing in compaction triggers, data layout, and data movement policies~\cite{sarkar2022constructing}. These differences primarily manifest in the prepare phase, where compaction is triggered and the input SSTables and merge key ranges are determined based on each strategy’s layout and movement policy. Once the input set is fixed, however, the subsequent run phase follows a common iterator-based execution flow: block-level reads, key comparison and merging, and new SSTable creation (Figure~\ref{fig:back_compaction}).
\sol{} is designed to accelerate the block read path that occurs repeatedly during this run phase. Therefore, it is applicable across leveled, tiered, and hybrid strategies, regardless of the compaction strategy. However, the frequency of system calls may vary depending on the fan-in scale or nesting depth of the input SSTables. Consequently, the magnitude of the acceleration effect may differ by strategy.

\vspace{-6pt}
\section{Conclusion}
\vspace{-4pt}
We proposed \sol{}, which integrates eBPF and io\_uring to migrate compaction I/O routines into the kernel, eliminating unnecessary user–kernel transitions and reducing software stack overhead. Experimental results with db\_bench and YCSB show that \sol{} reduces system call invocations during compaction by 99\% on average and shortens compaction time by 50\% compared to baseline RocksDB, improving throughput by up to 75\% and reducing p99 latency by 40\% under write-intensive workloads.

\bibliographystyle{ieeetr}
\bibliography{ref} 

@inproceedings{kim2025kvaccel,
  title={KVACCEL: A Novel Write Accelerator for LSM-Tree-Based KV Stores with Host-SSD Collaboration},
  author={Kim, Kihwan and Chung, Hyunsun and Ahn, Seonghoon and Park, Junhyeok and Jamil, Safdar and Byun, Hongsu and Lee, Myungcheol and Choi, Jinchun and Kim, Youngjae},
  booktitle={2025 IEEE International Parallel and Distributed Processing Symposium (IPDPS)},
  pages={666--677},
  year={2025},
  organization={IEEE}
}

@inproceedings{kim2025eco,
  title={ECO-KVS: Energy-Aware Compaction Offloading Mechanism for LSM-Tree Based Key-Value Stores in Edge Federation},
  author={Kim, Jeeseob and Byun, Hongsu and Lee, Seungjae and Kim, Myoungjoon and Kim, Youngjae and Xie, Zaipeng and Park, Sungyong},
  booktitle={2025 IEEE 25th International Symposium on Cluster, Cloud and Internet Computing (CCGrid)},
  pages={570--579},
  year={2025},
  organization={IEEE}
}

@inproceedings{kim2024coordinating,
  title={Coordinating compaction between lsm-tree based key-value stores for edge federation},
  author={Kim, Jeeseob and Yoo, Honghyeon and Lee, Seungjae and Byun, Hongsu and Park, Sungyong},
  booktitle={2024 IEEE 17th International Conference on Cloud Computing (CLOUD)},
  pages={419--429},
  year={2024},
  organization={IEEE}
}

@inproceedings{byun2025revisiting,
  title={Revisiting Multi-threaded Compaction in LSM-trees: Enabling Compaction Pipelining},
  author={Byun, Hongsu and Yoo, Honghyeon and Park, Sungyong},
  booktitle={Proceedings of the 54th International Conference on Parallel Processing},
  pages={794--803},
  year={2025}
}

@inproceedings{lepers2019kvell,
  title={Kvell: the design and implementation of a fast persistent key-value store},
  author={Lepers, Baptiste and Balmau, Oana and Gupta, Karan and Zwaenepoel, Willy},
  booktitle={Proceedings of the 27th ACM Symposium on Operating Systems Principles},
  pages={447--461},
  year={2019}
}

@article{matsunobu2020myrocks,
  title={Myrocks: Lsm-tree database storage engine serving facebook's social graph},
  author={Matsunobu, Yoshinori and Dong, Siying and Lee, Herman},
  journal={Proceedings of the VLDB Endowment},
  volume={13},
  number={12},
  pages={3217--3230},
  year={2020},
  publisher={VLDB Endowment}
}

@misc{sysbench,
  author = {Kopytov, Alexy},
  title = {sysbench},
  year = {2020},
  howpublished = {\url{https://github.com/akopytov/sysbench}}
}

@inproceedings{cooper2010benchmarking,
  title={Benchmarking cloud serving systems with YCSB},
  author={Cooper, Brian F and Silberstein, Adam and Tam, Erwin and Ramakrishnan, Raghu and Sears, Russell},
  booktitle={Proceedings of the 1st ACM symposium on Cloud computing},
  pages={143--154},
  year={2010}
}

@misc{Micron_7600_NVMe_2025,
  title        = {7600 NVMe SSD (PCIe Gen5) Product Brief},
  author       = {{Micron Technology}},
  howpublished = {\url{https://www.micron.com/products/storage/ssd/data-center-ssd/7600-ssd}},
  year         = {2025},
  note         = {Accessed: 2025-10-28; “low and consistent latency” stated.}
}

@misc{Intel_Optane_P5800X_2021,
  title        = {Optane P5800X SSD Product Brief},
  author       = {{Intel Corporation}},
  howpublished = {\url{https://www.techradar.com/computing/cloud-computing/an-obscure-4-year-old-intel-ssd-is-still-the-worlds-fastest}},
  year         = {2025},
  note         = {Accessed: 2025-10-28; latency “sub-10 microseconds” reported in review.}
}

@misc{SSSTC_EJ5_U2_2024,
  title        = {EJ5 U.2 (PCIe 5.0) Enterprise SSD},
  author       = {{SSSTC ( Solid State Storage Technology Corporation )}},
  howpublished = {\url{https://www.ssstc.com/products-detail/EJ5-U2-eTLC-pcie-ssd/}},
  year         = {2024},
  note         = {Accessed: 2025-10-28}
}

@misc{KIOXIA_CD8P_R_2023,
  title        = {CD8P-R Series (PCIe 5.0) Data Center NVMe™ SSDs},
  author       = {{KIOXIA Corporation}},
  howpublished = {\url{https://americas.kioxia.com/en-us/business/ssd/data-center-ssd/cd8p-r-e3s.html}},
  year         = {2023},
  note         = {Accessed: 2025-10-28}
}

@misc{Samsung_9100PRO_2025,
  title        = {9100 PRO Series SSD (PCIe 5.0) NVMe™},
  author       = {{Samsung Electronics}},
  howpublished = {\url{https://semiconductor.samsung.com/consumer-storage/internal-ssd/9100-pro/}},
  year         = {2025},
  note         = {Accessed: 2025-10-28}
}

@article{he2023database,
  title={When database meets new storage devices: Understanding and exposing performance mismatches via configurations},
  author={He, Haochen and Xu, Erci and Li, Shanshan and Jia, Zhouyang and Zheng, Si and Yu, Yue and Ma, Jun and Liao, Xiangke},
  journal={Proceedings of the VLDB Endowment},
  volume={16},
  number={7},
  pages={1712--1725},
  year={2023},
  publisher={VLDB Endowment}
}

@inproceedings{zhang2019m,
  title={I'm not dead yet! the role of the operating system in a kernel-bypass era},
  author={Zhang, Irene and Liu, Jing and Austin, Amanda and Roberts, Michael Lowell and Badam, Anirudh},
  booktitle={Proceedings of the Workshop on Hot Topics in Operating Systems},
  pages={73--80},
  year={2019}
}

@misc{dpdk,
  title        = {Data Plane Development Kit (DPDK)},
  author       = {{DPDK Project}},
  howpublished = {\url{https://www.dpdk.org/}},
  note         = {Accessed: 2025-10-28},
  year         = {2025}
}

@inproceedings{yang2017spdk,
  title={SPDK: A development kit to build high performance storage applications},
  author={Yang, Ziye and Harris, James R and Walker, Benjamin and Verkamp, Daniel and Liu, Changpeng and Chang, Cunyin and Cao, Gang and Stern, Jonathan and Verma, Vishal and Paul, Luse E},
  booktitle={2017 IEEE International Conference on Cloud Computing Technology and Science (CloudCom)},
  pages={154--161},
  year={2017},
  organization={IEEE}
}

@inproceedings{lee2019asynchronous,
  title={Asynchronous I/O stack: A low-latency kernel $\{$I/O$\}$ stack for Ultra-Low latency SSDs},
  author={Lee, Gyusun and Shin, Seokha and Song, Wonsuk and Ham, Tae Jun and Lee, Jae W and Jeong, Jinkyu},
  booktitle={2019 USENIX Annual Technical Conference (USENIX ATC 19)},
  pages={603--616},
  year={2019}
}

@inproceedings {196386,
author = {Hyeong-Jun Kim and Young-Sik Lee and Jin-Soo Kim},
title = {{NVMeDirect}: A User-space {I/O} Framework for Application-specific Optimization on {NVMe} {SSDs}},
booktitle = {8th USENIX Workshop on Hot Topics in Storage and File Systems (HotStorage 16)},
year = {2016},
address = {Denver, CO},
url = {https://www.usenix.org/conference/hotstorage16/workshop-program/presentation/kim},
publisher = {USENIX Association},
month = jun
}

@article{fleming2017_ebpf,
  author    = {Matt Fleming},
  title     = {A thorough introduction to eBPF},
  journal   = {LWN.net},
  year      = {2017},
  month     = {Dec},
  url       = {https://lwn.net/Articles/740157/},
  note      = {Accessed: 2025-10-28}
}

@misc{man7-iouring,
  author       = {Michael Kerrisk and others},
  title        = {io\_uring(7) --- Linux asynchronous I/O API},
  howpublished = {man7.org Linux man-pages},
  year         = {2024},
  url          = {\url{https://man7.org/linux/man-pages/man7/io_uring.7.html}}
}

@article{spdk,
    month = {February},
    author = {Karol Latecki and Maciej Wawryk},
    title = {{SPDK NVMe BDEV Performance Report Release 22.01}},
    type = {Technical Report},
    publisher = {Intel},
    year = {2022},
    pages = {11--12},
    url = {https://ci.spdk.io/download/performance-reports/SPDK\_nvme\_bdev\_perf\_report\_2201.pdf},
}

@article{lakshman2010cassandra,
  title={{Cassandra: A Decentralized Structured Storage System}},
  author={Lakshman, Avinash and Malik, Prashant},
  journal={ACM SIGOPS Operating Systems Review},
  volume={44},
  number={2},
  pages={35--40},
  year={2010},
  publisher={ACM New York, NY, USA}
}

@misc{levelDB,
    title={{LevelDB}},
    author={Google},
    howpublished={\url{https://github.com/google/leveldb}},
    year={2017}
}

@article{sarkar2022constructing,
  title={Constructing and analyzing the LSM compaction design space (updated version)},
  author={Sarkar, Subhadeep and Staratzis, Dimitris and Zhu, Zichen and Athanassoulis, Manos},
  journal={arXiv preprint arXiv:2202.04522},
  year={2022}
}

@misc{rocksdb-compaction,
  author       = {{Facebook}},
  title        = {Compaction · RocksDB},
  year         = {2023},
  howpublished = {\url{https://github.com/facebook/rocksdb/wiki/Compaction}},
  note         = {Accessed: 2025-05-01}
}

@article{o1996log,
  title={The log-structured merge-tree (LSM-tree)},
  author={O’Neil, Patrick and Cheng, Edward and Gawlick, Dieter and O’Neil, Elizabeth},
  journal={Acta Informatica},
  volume={33},
  pages={351--385},
  year={1996},
  publisher={Springer}
}

@misc{rocksdb,
  author       = {Facebook},
  title        = {RocksDB: A Persistent Key-Value Store for Fast Storage Environments},
  year         = {2013},
  howpublished = {\url{https://github.com/facebook/rocksdb}},
  note         = {Accessed: 2025-05-01}
}

@inproceedings{cao2020characterizing,
  title={Characterizing, modeling, and benchmarking RocksDB Key-Value workloads at facebook},
  author={Cao, Zhichao and Dong, Siying and Vemuri, Sagar and Du, David HC},
  booktitle={18th USENIX Conference on File and Storage Technologies (FAST 20)},
  pages={209--223},
  year={2020}
}

@inproceedings{liu2024understanding,
  title={Understanding performance of ebpf maps},
  author={Liu, Chang and Tak, Byungchul and Wang, Long},
  booktitle={Proceedings of the ACM SIGCOMM 2024 Workshop on eBPF and Kernel Extensions},
  pages={9--15},
  year={2024}
}

@inproceedings{volpert2025towards,
  title={Towards eBPF Overhead Quantification: An Exemplary Comparison of eBPF and SystemTap},
  author={Volpert, Simon and Winkelhofer, Sascha and Domaschka, J{\"o}rg and Wesner, Stefan},
  booktitle={Companion of the 16th ACM/SPEC International Conference on Performance Engineering},
  pages={122--129},
  year={2025}
}

@inproceedings{zhong2022xrp,
  title={{XRP:In-Kernel storage functions with eBPF}},
  author={Zhong, Yuhong and Li, Haoyu and Wu, Yu Jian and Zarkadas, Ioannis and Tao, Jeffrey and Mesterhazy, Evan and Makris, Michael and Yang, Junfeng and Tai, Amy and Stutsman, Ryan and others},
  booktitle={16th USENIX Symposium on Operating Systems Design and Implementation (OSDI 22)},
  pages={375--393},
  year={2022}
}

@online{dbbench,
  author        = {{Facebook}},
  title         = {{DB Bench}},
  url           = "https://github.com/facebook/rocksdb/wiki/Benchmarking-tools",
howpublished={\url{https://github.com/facebook/rocksdb/wiki/Benchmarking-tools}},
  year = {2017},
}

@online{writestall,
  author        = {{Facebook}},
  title         = {{Write Stalls}},
  url           = "https://github.com/facebook/rocksdb/wiki/Write-Stalls",
howpublished={\url{https://github.com/facebook/rocksdb/wiki/Write-Stalls}},
  year = {2021},
}

@inproceedings{wang2023rbc,
  title={{RBC: A bandwidth controller to reduce write-stalls and tail latency}},
  author={Wang, Zepeng and Yin, Shu},
  booktitle={Proceedings of the 52nd International Conference on Parallel Processing},
  pages={213--222},
  year={2023}
}

@inproceedings{yao2020matrixkv,
  title={{MatrixKV: Reducing write stalls and write amplification in LSM-tree based KV stores with matrix container in NVM}},
  author={Yao, Ting and Zhang, Yiwen and Wan, Jiguang and Cui, Qiu and Tang, Liu and Jiang, Hong and Xie, Changsheng and He, Xubin},
  booktitle={2020 USENIX Annual Technical Conference (USENIX ATC 20)},
  pages={17--31},
  year={2020}
}

@inproceedings{Dayan2018Dostoevsky,
    author = {Dayan, Niv and Idreos, Stratos},
    title = {Dostoevsky: Better Space-Time Trade-Offs for LSM-Tree Based Key-Value Stores via Adaptive Removal of Superfluous Merging},
    year = {2018},
    isbn = {9781450347037},
    publisher = {Association for Computing Machinery},
    address = {New York, NY, USA},
    url = {https://doi.org/10.1145/3183713.3196927},
    doi = {10.1145/3183713.3196927},
    booktitle = {Proceedings of the 2018 International Conference on Management of Data},
    pages = {505–520},
    numpages = {16},
    location = {Houston, TX, USA},
    series = {SIGMOD '18}
}

@article{Dayan2022Spooky,
    author = {Dayan, Niv and Weiss, Tamar and Dashevsky, Shmuel and Pan, Michael and Bortnikov, Edward and Twitto, Moshe},
    title = {Spooky: granulating LSM-tree compactions correctly},
    year = {2022},
    issue_date = {July 2022},
    publisher = {VLDB Endowment},
    volume = {15},
    number = {11},
    issn = {2150-8097},
    url = {https://doi.org/10.14778/3551793.3551853},
    doi = {10.14778/3551793.3551853},
    journal = {Proc. VLDB Endow.},
    month = jul,
    pages = {3071–3084},
    numpages = {14}
}

@inproceedings{Zhang2014Pipelined,
    author={Zhang, Zigang and Yue, Yinliang and He, Bingsheng and Xiong, Jin and Chen, Mingyu and Zhang, Lixin and Sun, Ninghui},
    booktitle={2014 IEEE 28th International Parallel and Distributed Processing Symposium}, 
    title={Pipelined Compaction for the LSM-Tree}, 
    year={2014},
    volume={},
    number={},
    pages={777-786},
    doi={10.1109/IPDPS.2014.85}
}

@article{Chai2020Adaptive,
    author={Chai, Yunpeng and Chai, Yanfeng and Wang, Xin and Wei, Haocheng and Wang, Yangyang},
    journal={IEEE Transactions on Knowledge and Data Engineering}, 
    title={Adaptive Lower-Level Driven Compaction to Optimize LSM-Tree Key-Value Stores}, 
    year={2022},
    volume={34},
    number={6},
    pages={2595-2609},
    doi={10.1109/TKDE.2020.3019264}
}

@article{Pan2017dCompaciton,
    author = {Pan, Fengfeng and Yue, Yinliang and Xiong, Jin},
    title = {dCompaction: Delayed Compaction for the LSM-Tree},
    year = {2017},
    issue_date = {December  2017},
    publisher = {Kluwer Academic Publishers},
    address = {USA},
    volume = {45},
    number = {6},
    issn = {0885-7458},
    url = {https://doi.org/10.1007/s10766-016-0472-z},
    doi = {10.1007/s10766-016-0472-z},
    journal = {Int. J. Parallel Program.},
    month = dec,
    pages = {1310–1325},
    numpages = {16}
}

@inproceedings{Sears2012blsm,
    author = {Sears, Russell and Ramakrishnan, Raghu},
    title = {bLSM: a general purpose log structured merge tree},
    year = {2012},
    isbn = {9781450312479},
    publisher = {Association for Computing Machinery},
    address = {New York, NY, USA},
    url = {https://doi.org/10.1145/2213836.2213862},
    doi = {10.1145/2213836.2213862},
    booktitle = {Proceedings of the 2012 ACM SIGMOD International Conference on Management of Data},
    pages = {217–228},
    numpages = {12},
    location = {Scottsdale, Arizona, USA},
    series = {SIGMOD '12}
}

@article{Teng2018Lsbmtree,
    author = {Teng, Dejun and Guo, Lei and Lee, Rubao and Chen, Feng and Zhang, Yanfeng and Ma, Siyuan and Zhang, Xiaodong},
    title = {A Low-cost Disk Solution Enabling LSM-tree to Achieve High Performance for Mixed Read/Write Workloads},
    year = {2018},
    issue_date = {May 2018},
    publisher = {Association for Computing Machinery},
    address = {New York, NY, USA},
    volume = {14},
    number = {2},
    issn = {1553-3077},
    url = {https://doi.org/10.1145/3162615},
    doi = {10.1145/3162615},
    journal = {ACM Trans. Storage},
    month = apr,
    articleno = {15},
    numpages = {26}
}

@inproceedings {Wu2015Lsmtrie,
    author = {Xingbo Wu and Yuehai Xu and Zili Shao and Song Jiang},
    title = {LSM-trie: An LSM-tree-based Ultra-Large Key-Value Store for Small Data Items},
    booktitle = {2015 USENIX Annual Technical Conference (USENIX ATC 15)},
    year = {2015},
    isbn = {978-1-931971-225},
    address = {Santa Clara, CA},
    pages = {71--82},
    url = {https://www.usenix.org/conference/atc15/technical-session/presentation/wu},
    publisher = {USENIX Association},
    month = jul
}

@inproceedings{Wang2022Blockgrained,
    author={Wang, Xiaoliang and Jin, Peiquan and Hua, Bei and Long, Hai and Huang, Wei},
    booktitle={2022 IEEE 38th International Conference on Data Engineering (ICDE)}, 
    title={Reducing Write Amplification of LSM-Tree with Block-Grained Compaction}, 
    year={2022},
    volume={},
    number={},
    pages={3119-3131},
    doi={10.1109/ICDE53745.2022.00279}
}

@article{Sun2023Finegrained,
    author={Sun, Hui and Chen, Guanzhong and Yue, Yinliang and Qin, Xiao},
    journal={IEEE Transactions on Cloud Computing}, 
    title={Improving LSM-Tree Based Key-Value Stores With Fine-Grained Compaction Mechanism}, 
    year={2023},
    volume={11},
    number={4},
    pages={3778-3796},
    doi={10.1109/TCC.2023.3329646}
}

@article{ding2022trianglekv,
    author={Ding, Chen and Yao, Ting and Jiang, Hong and Cui, Qiu and Tang, Liu and Zhang, Yiwen and Wan, Jiguang and Tan, Zhihu},
    journal={IEEE Transactions on Parallel and Distributed Systems}, 
    title={TriangleKV: Reducing Write Stalls and Write Amplification in LSM-Tree Based KV Stores With Triangle Container in NVM}, 
    year={2022},
    volume={33},
    number={12},
    pages={4339-4352},
    doi={10.1109/TPDS.2022.3188268}
}

@inproceedings {Balmau2019SILK,
    author = {Oana Balmau and Florin Dinu and Willy Zwaenepoel and Karan Gupta and Ravishankar Chandhiramoorthi and Diego Didona},
    title = {{SILK}: Preventing Latency Spikes in {Log-Structured} Merge {Key-Value} Stores},
    booktitle = {2019 USENIX Annual Technical Conference (USENIX ATC 19)},
    year = {2019},
    isbn = {978-1-939133-03-8},
    address = {Renton, WA},
    pages = {753--766},
    url = {https://www.usenix.org/conference/atc19/presentation/balmau},
    publisher = {USENIX Association},
    month = jul
}

@inproceedings{Jin2021DLC,
    title={DLC: A New Compaction Scheme for LSM-tree with High Stability and Low Latency},
    author={Peiquan Jin and Jianchuan Li and Hai Long},
    booktitle={International Conference on Extending Database Technology},
    year={2021},
    url={https://api.semanticscholar.org/CorpusID:232283984}
}

@inproceedings{Yu2023ADOC,
    author = {Jinghuan Yu and Sam H. Noh and Young-ri Choi and Chun Jason Xue},
    title = {ADOC: Automatically Harmonizing Dataflow Between Components in Log-Structured Key-Value Stores for Improved Performance},
    booktitle = {21st USENIX Conference on File and Storage Technologies (FAST 23)},
    year = {2023},
    isbn = {978-1-939133-32-8},
    address = {Santa Clara, CA},
    pages = {65--80},
    url = {https://www.usenix.org/conference/fast23/presentation/yu},
    publisher = {USENIX Association},
    month = feb
}

@article{Zhang2024Succintkv,
    author = {Zhang, Yinan and Yang, Shun and Hu, Huiqi and Yang, Chengcheng and Cai, Peng and Zhou, Xuan},
    title = {SuccinctKV: a CPU-efficient LSM-tree Based KV Store with Scan-based Compaction},
    year = {2024},
    issue_date = {December 2024},
    publisher = {Association for Computing Machinery},
    address = {New York, NY, USA},
    volume = {21},
    number = {4},
    issn = {1544-3566},
    url = {https://doi.org/10.1145/3695873},
    doi = {10.1145/3695873},
    journal = {ACM Trans. Archit. Code Optim.},
    month = nov,
    articleno = {90},
    numpages = {26}
}

@inproceedings{Zhao2021Wipdb,
    author={Zhao, Xingsheng and Jiang, Song and Wu, Xingbo},
    booktitle={2021 IEEE 37th International Conference on Data Engineering (ICDE)}, 
    title={WipDB: A Write-in-place Key-value Store that Mimics Bucket Sort}, 
    year={2021},
    volume={},
    number={},
    pages={1404-1415},
    doi={10.1109/ICDE51399.2021.00125}
}

@inproceedings{Raju2017PebblesDB,
    author = {Raju, Pandian and Kadekodi, Rohan and Chidambaram, Vijay and Abraham, Ittai},
    title = {PebblesDB: Building Key-Value Stores using Fragmented Log-Structured Merge Trees},
    year = {2017},
    isbn = {9781450350853},
    publisher = {Association for Computing Machinery},
    address = {New York, NY, USA},
    url = {https://doi.org/10.1145/3132747.3132765},
    doi = {10.1145/3132747.3132765},
    booktitle = {Proceedings of the 26th Symposium on Operating Systems Principles},
    pages = {497–514},
    numpages = {18},
    location = {Shanghai, China},
    series = {SOSP '17}
}

@inproceedings{zhang2020fpgacompaction,
  author    = {Teng Zhang and others},
  title     = {FPGA-Accelerated Compactions for LSM-based Key-Value Stores},
  booktitle = {18th USENIX Conference on File and Storage Technologies (FAST ’20)},
  year      = {2020},
  pages     = {183–198},
  publisher = {USENIX Association},
  note      = {Also appears in USENIX FAST ’20}
}

@inproceedings{Sun2020FPGA,
  author={Sun, Xuan and Yu, Jinghuan and Zhou, Zimeng and Xue, Chun Jason},
  booktitle={2020 IEEE 36th International Conference on Data Engineering (ICDE)}, 
  title={FPGA-based Compaction Engine for Accelerating LSM-tree Key-Value Stores}, 
  year={2020},
  volume={},
  number={},
  pages={1261-1272},
  doi={10.1109/ICDE48307.2020.00113}}

@inproceedings{Zhou2024GPUCompaction,
  author    = {Hao Zhou and others},
  title     = {A GPU-Accelerated Compaction Strategy for LSM-based Key-Value Store},
  booktitle = {Proceedings of the 2024 MSST Conference},
  year      = {2024},
  note      = {Implemented on LevelDB, reports up to 3.61× compaction speed-up.}
}

@article{Sun2024glsm,
    author = {Sun, Hui and Xu, Jinfeng and Jiang, Xiangxiang and Chen, Guanzhong and Yue, Yinliang and Qin, Xiao},
    title = {gLSM: Using GPGPU to Accelerate Compactions in LSM-tree-based Key-value Stores},
    year = {2024},
    issue_date = {February 2024},
    publisher = {Association for Computing Machinery},
    address = {New York, NY, USA},
    volume = {20},
    number = {1},
    issn = {1553-3077},
    url = {https://doi.org/10.1145/3633782},
    doi = {10.1145/3633782},
    journal = {ACM Trans. Storage},
    month = jan,
    articleno = {5},
    numpages = {41}
}

@article{Xu2020LUDA,
    author    = {Peng Xu and Jiguang Wan and Ping Huang and Xiaogang Yang and Chenlei Tang and Fei Wu and Changsheng Xie},
    title     = {LUDA: Boost LSM Key Value Store Compactions with GPUs},
    journal   = {arXiv preprint arXiv:2004.03054},
    year      = {2020},
    note      = {Preprint}
}

@article{Pan2017dCompaction,
    author    = {Feng-Feng Pan and Yin-Liang Yue and Jin Xiong},
    title     = {dCompaction: Speeding up Compaction of the LSM-Tree via Delayed Compaction},
    journal   = {Journal of Computer Science and Technology},
    year      = {2017},
    volume    = {32},
    number    = {1},
    pages     = {41–54},
    doi       = {10.1007/s11390-017-1704-4}
}

@article{Sun2019NDP,
author = {Sun, Hui and Liu, Wei and Huang, Jianzhong and Shi, Weisong},
title = {Collaborative Compaction Optimization System using Near-Data Processing for LSM-tree-based Key-Value Stores},
year = {2019},
issue_date = {Sep 2019},
publisher = {Academic Press, Inc.},
address = {USA},
volume = {131},
number = {C},
issn = {0743-7315},
url = {https://doi.org/10.1016/j.jpdc.2019.04.011},
doi = {10.1016/j.jpdc.2019.04.011},
journal = {J. Parallel Distrib. Comput.},
month = sep,
pages = {29–43},
numpages = {15}
}

@article{Sun2024NDP,
    author = {Sun, Hui and Lou, Bendong and Zhao, Chao and Kong, Deyan and Zhang, Chaowei and Huang, Jianzhong and Yue, Yinliang and Qin, Xiao},
    title = {Asynchronous Compaction Acceleration Scheme for Near-data Processing-enabled LSM-tree-based KV Stores},
    year = {2024},
    issue_date = {November 2024},
    publisher = {Association for Computing Machinery},
    address = {New York, NY, USA},
    volume = {23},
    number = {6},
    issn = {1539-9087},
    url = {https://doi.org/10.1145/3626097},
    doi = {10.1145/3626097},
    journal = {ACM Trans. Embed. Comput. Syst.},
    month = sep,
    articleno = {93},
    numpages = {33}
}

@inproceedings{Li2022FaaS,
author = {Li, Jianchuan and Jin, Peiquan and Lin, Yuanjin and Zhao, Ming and Wang, Yi and Guo, Kuankuan},
title = {Elastic and Stable Compaction for LSM-tree: A FaaS-Based Approach on TerarkDB},
year = {2021},
isbn = {9781450384469},
publisher = {Association for Computing Machinery},
address = {New York, NY, USA},
url = {https://doi.org/10.1145/3459637.3481913},
doi = {10.1145/3459637.3481913},
pages = {3906–3915},
numpages = {10},
location = {Virtual Event, Queensland, Australia},
series = {CIKM '21}
}

@article{Yu2024CaaS,
author = {Yu, Qiaolin and Guo, Chang and Zhuang, Jay and Thakkar, Viraj and Wang, Jianguo and Cao, Zhichao},
title = {CaaS-LSM: Compaction-as-a-Service for LSM-based Key-Value Stores in Storage Disaggregated Infrastructure},
year = {2024},
issue_date = {June 2024},
publisher = {Association for Computing Machinery},
address = {New York, NY, USA},
volume = {2},
number = {3},
url = {https://doi.org/10.1145/3654927},
doi = {10.1145/3654927},
journal = {Proc. ACM Manag. Data},
month = may,
articleno = {124},
numpages = {28}
}

@article{Lu2016WiscKey,
author = {Lu, Lanyue and Pillai, Thanumalayan Sankaranarayana and Gopalakrishnan, Hariharan and Arpaci-Dusseau, Andrea C. and Arpaci-Dusseau, Remzi H.},
title = {WiscKey: Separating Keys from Values in SSD-Conscious Storage},
year = {2017},
issue_date = {February 2017},
publisher = {Association for Computing Machinery},
address = {New York, NY, USA},
volume = {13},
number = {1},
issn = {1553-3077},
url = {https://doi.org/10.1145/3033273},
doi = {10.1145/3033273},
journal = {ACM Trans. Storage},
month = mar,
articleno = {5},
numpages = {28}
}

@inproceedings {Helen2018HashKV,
author = {Helen H. W. Chan and Yongkun Li and Patrick P. C. Lee and Yinlong Xu},
title = {{HashKV}: Enabling Efficient Updates in {KV} Storage via Hashing},
booktitle = {2018 USENIX Annual Technical Conference (USENIX ATC 18)},
year = {2018},
isbn = {978-1-931971-44-7},
address = {Boston, MA},
pages = {1007--1019},
url = {https://www.usenix.org/conference/atc18/presentation/chan},
publisher = {USENIX Association},
month = jul
}

@inproceedings {Li2021DiffKV,
author = {Yongkun Li and Zhen Liu and Patrick P. C. Lee and Jiayu Wu and Yinlong Xu and Yi Wu and Liu Tang and Qi Liu and Qiu Cui},
title = {Differentiated {Key-Value} Storage Management for Balanced {I/O} Performance},
booktitle = {2021 USENIX Annual Technical Conference (USENIX ATC 21)},
year = {2021},
isbn = {978-1-939133-23-6},
pages = {673--687},
url = {https://www.usenix.org/conference/atc21/presentation/li-yongkun},
publisher = {USENIX Association},
month = jul
}

@inproceedings {Zhou2023Userspacebypass,
author = {Zhe Zhou and Yanxiang Bi and Junpeng Wan and Yangfan Zhou and Zhou Li},
title = {Userspace Bypass: Accelerating Syscall-intensive Applications},
booktitle = {17th USENIX Symposium on Operating Systems Design and Implementation (OSDI 23)},
year = {2023},
isbn = {978-1-939133-34-2},
address = {Boston, MA},
pages = {33--49},
url = {https://www.usenix.org/conference/osdi23/presentation/zhou-zhe},
publisher = {USENIX Association},
month = jul
}

@inproceedings{Zhssman2025cacheext,
author = {Zussman, Tal and Zarkadas, Ioannis and Carin, Jeremy and Cheng, Andrew and Franke, Hubertus and Pfefferle, Jonas and Cidon, Asaf},
title = {cache\_ext: Customizing the Page Cache with eBPF},
year = {2025},
isbn = {9798400718700},
publisher = {Association for Computing Machinery},
address = {New York, NY, USA},
url = {https://doi.org/10.1145/3731569.3764820},
doi = {10.1145/3731569.3764820},
booktitle = {Proceedings of the ACM SIGOPS 31st Symposium on Operating Systems Principles},
pages = {462–478},
numpages = {17},
location = {Lotte Hotel World, Seoul, Republic of Korea},
series = {SOSP '25}
}

@inproceedings{Zussman2024Custom,
author = {Zussman, Tal and Jiang, Teng and Cidon, Asaf},
title = {Custom Page Fault Handling With eBPF},
year = {2024},
isbn = {9798400707124},
publisher = {Association for Computing Machinery},
address = {New York, NY, USA},
url = {https://doi.org/10.1145/3672197.3673432},
doi = {10.1145/3672197.3673432},
booktitle = {Proceedings of the ACM SIGCOMM 2024 Workshop on EBPF and Kernel Extensions},
pages = {71–73},
numpages = {3},
location = {Sydney, NSW, Australia},
series = {eBPF '24}
}

@article{butrovich2025bpf,
  title={{BPF-DB: A Kernel-Embedded Transactional Database Management System For eBPF Applications}},
  author={Butrovich, Matthew and Arch, Samuel and Lim, Wan Shen and Zhang, William and Patel, Jignesh M and Pavlo, Andrew},
  journal={Proceedings of the ACM on Management of Data},
  volume={3},
  number={3},
  pages={1--27},
  year={2025},
  publisher={ACM New York, NY, USA}
}

\end{document}